\documentclass[11pt]{article}

\usepackage[hypertexnames=false]{hyperref}
\usepackage[margin={2cm,2cm}]{geometry} 

\usepackage{sectsty}
\usepackage{abstract}
\usepackage{color}
\usepackage{graphicx}
\usepackage{multicol}
\usepackage{amsmath,amssymb,amstext}
\usepackage[usenames,svgnames]{xcolor}
\usepackage{subfigure}
\usepackage{authblk}
\usepackage{titlesec}
\usepackage{setspace}
\usepackage{times,fancyhdr}
\usepackage[normalem]{ulem} 
\usepackage{enumerate} 
\usepackage{mathrsfs} 
\usepackage{bigints}
\usepackage{bm}
\input{undertilde} 
\usepackage{mathtools, cuted}

\titlespacing*{\section}{0pt}{3ex plus 2ex}{1ex} 
\sectionfont{\fontsize{11}{11}\selectfont} 
\subsectionfont{\fontsize{10}{10}\selectfont} 

\hypersetup{
    colorlinks=true,
    urlcolor=SteelBlue,
    linkcolor=red,
    citecolor=blue,
}

\newcommand*{\Scale}[2][4]{\scalebox{#1}{$#2$}} 
\newcommand{\romansubs}{\renewcommand{\theequation}{\theparentequation \roman{equation}}} 
\newcommand{\piA}[2]{\mbox{\tiny{$_{A}$}}\tilde{\pi}_{#1}^{\;#2}}
\newcommand{\piE}[2]{\mbox{\tiny{$_{E}$}}\tilde{\pi}^{#1}_{\;#2}}
\newcommand{\dform}[1]{\bm{\mathrm{#1}}}
\newcommand{\te}{\tilde{\dform{e}}}
\newcommand{\tw}{\tilde{\dform{\omega}}}
\newcommand{\starf}{{^{*}\dform{F}}}
\newcommand{\starftilde}{{^{*}\tilde{\dform{F}}}}
\newcommand{\lie}{\text{\pounds}}

\begin{document}
\pagestyle{fancy}
\fancyhead{} 
\fancyhead[OR]{\thepage}
\fancyhead[OC]{{\small{GRAVITATIONAL CONSTRAINT ALGEBRA UNDER DEFORMATIONS}}}
\fancyfoot{} 
\renewcommand\headrulewidth{0.5pt}
\addtolength{\headheight}{2pt} 
\global\long\def\tdud#1#2#3#4#5{#1_{#2}{}^{#3}{}_{#4}{}^{#5}}
\global\long\def\tudu#1#2#3#4#5{#1^{#2}{}_{#3}{}^{#4}{}_{#5}}

\twocolumn
\title{\bf{\normalsize{Hamiltonian consistency of the gravitational constraint algebra under deformations}}}


\author[1,2]{\small{J. T. G\'{a}lvez Ghersi}\thanks{\href{mailto:joseg@sfu.ca}{joseg@sfu.ca}}}%

\author[3]{\small{M. J. Desrochers}\thanks{\href{mailto:mjdesrochers@phas.ubc.ca}{mjdesrochers@phas.ubc.ca}}}%

\author[4,1]{\small{M. Protter}\thanks{\href{mailto:protter@ualberta.ca}{protter@ualberta.ca}}}%

\author[5,1]{\small{A. DeBenedictis}\thanks{\href{mailto:adebened@sfu.ca}{adebened@sfu.ca}}}%

\affil[1]{\footnotesize{\it{Department of Physics, Simon Fraser University}}\\
\footnotesize{\it{8888 University Drive, Burnaby, BC, V5A 1S6, Canada}}}
\affil[2]{\footnotesize{\it{Perimeter Institute for Theoretical Physics}}\\
\footnotesize{\it{31 Caroline Street North, Waterloo, ON, N2L 2Y5, Canada}}}
\affil[3]{\footnotesize{\it{Department of Physics and Astronomy, The University of British Columbia}}\\
\footnotesize{\it{6224 Agricultural Road, Vancouver, BC, V6T 1Z1, Canada}}}
\affil[4]{\footnotesize{\it{Department of Physics, The University of Alberta}}\\
\footnotesize{\it{Edmonton, AB, T6G 2E1, Canada}}}
\affil[5]{\footnotesize{\it{The Pacific Institute for the Mathematical Sciences}}}

\date{\vspace{-0.8cm}(\footnotesize{\today})} 
\twocolumn[ 
  \begin{@twocolumnfalse}  
  \begin{changemargin}{1.75cm}{1.75cm} 
\maketitle
\end{changemargin} 
\vspace{-1.0cm}
\begin{changemargin}{1.5cm}{1.5cm} 
\begin{abstract}
{\noindent\small{The importance of the first-class constraint algebra of general relativity is not limited just by its self-contained description of the gauge nature of spacetime, but it also provides conditions to properly evolve the geometry by selecting a gauge only once throughout the whole evolution of a gravitational system. This must be a property of all background independent theories. In this paper we consider gravitational theories which arise from deformations of the fundamental canonical variables of general relativity where the proposed deformations are inspired by modifications of gravity. These variable deformations result in new theories when the deformation is not a canonical transformation. The new theory must preserve the first-class structure of the algebra, which is a non-trivial restriction for generic deformations. In this vein we present a general deformation scheme along with consistency conditions, so that the algebra of constraints is still satisfied in the resulting theory. This is illustrated both in metric theory as well as in tetrad theory.}}
\end{abstract}
\noindent{\footnotesize PACS(2010): 04.20.Fy\;\; 02.20.Sv\;\; 04.50.Kd\;;\; MSC(2010): 37K65\;\;70H40/5\;\;83C99\;\;83D05}
\rule{\linewidth}{0.2mm}
\end{changemargin}
\end{@twocolumnfalse} 
]
\saythanks 
\vspace{0.5cm}
{\setstretch{0.9} 
\section{Introduction}
Einstein's theory of general relativity (GR) is considered by many to be the pinnacle of classical field theories. It provides a powerful description of strong gravitational phenomena at solar system scales and its validity has been tested in a variety of local experiments discussed in \cite{Will:2014kxa, Abbott:2016blz}.  Nevertheless, there are reasonable arguments to propose modifications to it, such as the failure to produce a sensible interacting picture of quantum gravity at ultraviolet scales just from promoting the classical theory, or the intriguing nature of Dark Energy and Dark Matter, which emerge as a necessary component to explain the dynamics of the universe at cosmological scales. Regardless of the approach followed to modify Einstein's theory, the outcome should be consistent with the actual observations and must not be in conflict with the behavior of matter at scales where the standard model has accurate results, as mentioned in \cite{Burgess:2013ara}. 

On the other hand, the Hamiltonian form of all the degrees of freedom in general relativity -{}- which in the case we study will also include matter minimally coupled to gravity -{}- reveals its nature as a first-class constrained system. The total scalar and diffeomorphism constraints (and also the Gauss constraint in the case of Ashtekar variables) form a closed ``algebra'' with spacetime dependent structure constants. This is a valuable feature of any gravitational system for many reasons explored in \cite{Hojman:1976vp, Rovelli:1986hp}; namely it fixes the surface where gauge orbits lie. This means that in order to evolve spacetime one only needs to fix the coordinates once at the surface of initial conditions. It also reveals that time and gauge evolution of each component of the Hamiltonian follows the rules of Lie transport. Moreover, once the gauge is fixed, the closure of the algebra allows dynamical classical solutions that preserve diffeomorphism invariance without imposing further conditions at each instant of time. Therefore, it is plausible to require that the same constraint structure appears in a deformed theory, as this new theory should also respect diffeomorphism symmetry. Hence, in this paper we study the modifications of a theory of gravity starting from its canonical variables, in the context of Hamiltonian systems where time evolution can be separated from gauge orbits as proposed by Dirac in \cite{Dirac:1951zz}, in order to preserve the closed form of the constraint algebra. These deformations are relevant, for example, in the growing interest in numerical simulations of astrophysical objects in the context of deformed theories of gravity, or in cases where corrections to the original theory manifest themselves as deformations of the canonical variables. Specific areas where variable deformation is commonly done are, for example, in effective minisuperspace loop quantum gravity (where one applies holonomy corrections to the connection variable), as in \cite{Tibrewala:2013kba, Campiglia:2016fzp, Bojowald:2015zha}). In a slightly different vein, more recently a theory of gravity has been created by demanding that a general globally Lorentz invariant theory be promoted to local invariance, yielding a gravitational field theory as in \cite{ref:gauge1, ref:gauge2}.

Admittedly, one does not necessarily need to deform one theory in order to get another theory, and one could simply consider some new theory from scratch. However, as mentioned above, deformations of the type studied here are common, as often one wishes for the new theory to be related in some way to the original, nondeformed theory. In addition to this, some models of modified gravity with a closed constraint algebra will also arise from generic deformations of the canonical variables of the system, as suggested in \cite{Freidel:2008ku, Clifton:2011jh, Sebastiani:2016ras}. Even when in general these transformations are not canonical, which implies that these might introduce extra degrees of freedom in the system, notice that it is still possible to describe the system using Hamiltonian mechanics in terms of an expanded set of canonical variables in the new theory. In all the cases, we will derive these quantities by keeping the original fields as configuration variables, and calculating their corresponding new conjugate momenta. -{}- normally these will not coincide with the corrected momenta -{}-  which  are constrained by the way the corrected variables depend on the original ones. One of the motivations of these transformations is the usual modification of field variables after quantum corrections since in general, the corrected fields when replaced directly into the original action do not necessarily become new canonical variables of the system. Other field redefinitions occurring in certain theories of modified gravity have a similar effect. In a certain sense, we roughly explore the classical analog to the ``inverse'' transformation that integrates out degrees of freedom in order to obtain an effective field theory. Moreover, we provide conditions for those transformations in order to deform general relativity into another gauge theory of gravity. 

The plan of this paper is as follows: In section \ref{sec:motivation}, we introduce the type of transformations that deform the action of a theory and their Hamiltonian analogues. To illustrate this transformations, we first provide simple examples of one dimensional cases. In section \ref{sec:algebra}, we briefly review the Hamiltonian formalism of general relativity as our undeformed starting point, and the derivation of the constraint algebra from the gauge algebra. We will also discuss the main properties we should preserve after deformations. In section \ref{sec:GR}, we apply these transformations using the standard canonical variables of the ADM formalism for the Einstein-Hilbert action. Two possible scenarios will be presented: (1) the transformation has to be canonical in agreement with the Lovelock's theorem or (2) we introduce more degrees of freedom in the theory. In section \ref{sec:Ashtekar}, a transformation of the Ashtekar version of GR is performed in order to obtain either one of the many different theories of general relativity or a modified theory of gravity with extra degrees of freedom. Finally, we present our discussions and conclude.     
 
\section{Transforming the action from its canonical variables}\label{sec:motivation}
In this section, we present two different ways to deform an action of via the change of its canonical variables. To introduce these transformations several toy models are first examined as a segue to the much more complicated arena of gravitational field theory which follows. Let us consider the case of a 1-D time dependent system with an action
\begin{equation}
S_1=\int \mathcal{L}(q,\dot{q})dt,\label{1Daction}
\end{equation} 
where $q$ and $\dot{q}$ are the dynamical variables of the system. In both cases studied in the subsections below, different transformations of the action are performed via deforming its canonical variables. We find the corresponding Hamiltonian representation of the system by a Legendre transformation 
\begin{equation}
H(p,q)=p\dot{q}-\mathcal{L}(p,q),\nonumber
\end{equation} 
where $p=\partial\mathcal{L}/\partial\dot{q}$ is the conjugate momentum of the canonical variable $q$. Other cases in which higher-order derivative terms cancel out in a manner that generates second-order equations of motion, such as seen in \cite{Deffayet:2011gz} will not be considered.

\subsection{Transformations into theories with second-order equations of motion}\label{subsec:transI}
The purpose of this section is to show transformations that lead us to describe the dynamics of the system by second-order differential equations. The canonical variables $q$ and $\dot{q}$ are transformed in the following way 
\begin{equation}
 q\rightarrow Q(q,\dot{q})\:,\:\dot{q}\rightarrow \tilde{Q}(q,\dot{q})
\end{equation}
where both $Q$ and $\tilde{Q}$ do not introduce new derivatives of $q$ in the Lagrangian. In this case, the new action reads as
\begin{equation}
S_2=\int \mathcal{L}'\left(Q(q,\dot{q}),\tilde{Q}(q,\dot{q})\right)dt\,,\nonumber
\end{equation} 
where the prime denotes that after the substitution, since the configuration variable is to remain as $q$, the resulting Lagrangian is different the the original one. If the system remains integrability after these transformations -{}- which is always true in the case of a canonical transformation, since in that case $S_1=S_2$ -{}- it is possible to find the Hamiltonian version of both actions via invertible Legendre transformations. Each system can be mapped into the other as we describe in Fig.~\ref{fig:figureI}. 

\begin{figure}[!h]
\vspace{.5cm}
\begin{center}
\includegraphics[width=0.33\textwidth]{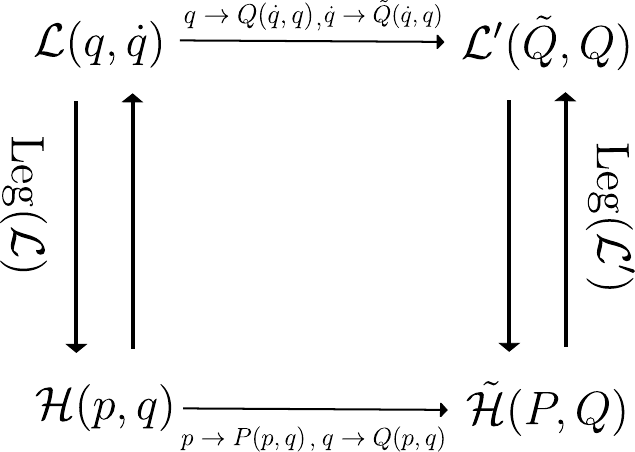}
\caption{\small{Mapping between $S_1$ and $S_2$ in its Lagrangian and Hamiltonian versions.}}
\label{fig:figureI}
\end{center}
\end{figure}

As a relevant example of this type of transformations, we will consider the case of a simple harmonic oscillator:
\begin{equation}
S_A=\frac{1}{2}\int \left(\dot{q}^2-\omega^2 q^2\right) dt,\label{1DSHO}
\end{equation} 
where the Hamiltonian is simply given by $H_A(p,q)=1/2\left(p^2+\omega^2q^2\right)$. We can write the same Hamiltonian via an arbitrary similarity transformation 
\begin{align}
&H_A=\frac{1}{2}\left[\begin{array}{cc} \mathcal{P} & \mathcal{Q}\end{array}\right]\left[\begin{array}{cc}\alpha & \beta\\
\beta & \gamma\omega^2\end{array}\right]\left[\begin{array}{c} \mathcal{P}\\ \mathcal{Q}\end{array}\right]\nonumber\\
&=\frac{1}{2}\left[\alpha\mathcal{P}^2+\gamma\omega^2\mathcal{Q}^2+2\beta\mathcal{P}\mathcal{Q}\right]\label{1DHSHO}
\end{align} 
where $\alpha$, $\beta$ and $\gamma$ are used as rotation parameters in phase space. In this example, it is easy to see that the similarity transformation to the auxiliary variables $\mathcal{P}$ and $\mathcal{Q}$ is canonical. These variables are related to the standard $p$ and $q$ by the characteristic orthonormal matrices of the similarity transformation. The off-diagonal term can be used to define a theory with a special class of solutions, to accomplish this one only need to define a function $f(t)$ such that 
\begin{equation}
\frac{1}{f}\frac{df}{dt}=2\beta,\nonumber
\end{equation}
and with the help of a new variable $\varphi=\mathcal{Q}/\sqrt{f}$, this off-diagonal term can be used as a generating function of canonical transformations
\begin{equation}
\frac{\partial G(\mathcal{P},\varphi)}{\partial t}=\frac{1}{\sqrt{f}}\frac{df}{dt}\mathcal{P}\varphi\:\rightarrow\:G(\mathcal{P},\varphi)=\sqrt{f}\mathcal{P}\varphi.\nonumber
\end{equation}
Therefore the Hamiltonian transforms via $H_A(\mathcal{P},\varphi)=H'_A(\mathcal{P},\varphi)+\partial G(\mathcal{P},\varphi)/\partial t$, the conjugate momentum of $\varphi$ is $\pi_{\varphi}=\partial G(\mathcal{P},\varphi)/\partial \mathcal{\varphi}=\sqrt{f}\mathcal{P}$, hence the Hamiltonian now ``drains'' energy from the kinetic term and adds it to the potential term via  
\begin{equation}
H'_A(\pi_\varphi,\varphi)=\frac{1}{2}\left[\alpha\frac{\pi_\varphi^2}{f}+f\gamma\omega^2\varphi^2\right],\label{DefH}
\end{equation}
and its corresponding Lagrangian is
\begin{equation}
\mathcal{L}'(\dot{\varphi},\varphi)=\frac{e^{2\beta t}}{2}\left[\alpha\dot{\varphi}^2-\gamma\omega^2\varphi^2\right]\label{DefL}.
\end{equation}
This remains a one-dimensional problem while preserving the number of degrees of freedom, but it adds a time-dependent scale factor similar to the case of an oscillator in an expanding geometry. To close the maps, it is possible to transform the Lagrangian \eqref{1DSHO} directly into \eqref{DefL} by deforming $\dot{q}\rightarrow\sqrt{\alpha}e^{\beta t}\dot{\varphi}$ and $q\rightarrow\sqrt{\gamma}e^{\beta t}\varphi$. In a similar way, we find the corresponding deformation of the Hamiltonian \eqref{1DHSHO} into \eqref{DefH} from $p\rightarrow\sqrt{\alpha}e^{-\beta t}\pi_{\varphi}$ and $q\rightarrow\sqrt{\gamma}e^{\beta t}\varphi$. One should note that it is only in rare circumstances that such a transformation is canonical and/or preserves the number of degrees of freedom in the system. It is convenient to write the deformations with respect to the original canonical variables in order to see the effect of the extra terms introduced in the new theory. The transformations in the Lagrangian can be written as deformations of the canonical variables
\begin{equation}
\dot{q}\rightarrow\dot{\varphi}+\Delta\dot{\varphi}\:,\: q\rightarrow\varphi+\Delta\varphi,\label{defcanonicaL}
\end{equation}  
and in an analogous way for the Hamiltonian
\begin{equation}
q\rightarrow\varphi+\Delta\varphi\:,\: p\rightarrow\pi_\varphi+\Delta\pi_\varphi,\label{defcanonicaH}
\end{equation}  
where we do not consider any specific range of magnitudes for $\Delta\dot{\varphi}$, $\Delta\varphi$ and $\Delta\pi_\varphi$ when compared with the canonical variables, although in many cases of interest these quantities can be obtained by any perturbative expansion of the original choice for a deformation. This decomposition of the deformed variables will be used in the remaining sections of this manuscript. Therefore, it is relevant to notice that the only difference between $\varphi$ and $q$ (including the canonical momenta $p$ and $\pi_{\varphi}$) is either a phase or a symmetry transformation that does not represent a significant modification of the role of the field variables. 

\subsection{Transformations into higher-order theories}\label{subsec:transII}
In this section, we consider another introductory example where we will promote a system with second-order equations of motion into a higher derivative theory by a different type of variable transformation:
\begin{equation}
q\rightarrow Q(q,\dot{q})\:,\:\dot{q}\rightarrow \frac{d}{dt}Q(q,\dot{q}).\label{transII}
\end{equation}   
After replacing in \eqref{1Daction}, the new action now reads
\begin{equation}
S_2=\int dt\:\mathcal{L}'\left(q, \dot{q},\frac{\partial Q}{\partial \dot{q}}\ddot{q}\right),
\label{1DacttransII}
\end{equation}
which is a function of the second derivative of the field. From the definition in \eqref{transII}, we can observe that it is enough to deform the field variable to promote the action since the derivative of the transformed field raises the order of the system. In analogy with the mapping proposed in the previous subsection \ref{subsec:transI}, we illustrate the mapping between $S_1$ and $S_2$ in Fig.~\ref{fig:figureII}. 

\begin{figure}[!h]
\vspace{.5cm}
\begin{center}
\includegraphics[width=0.37\textwidth]{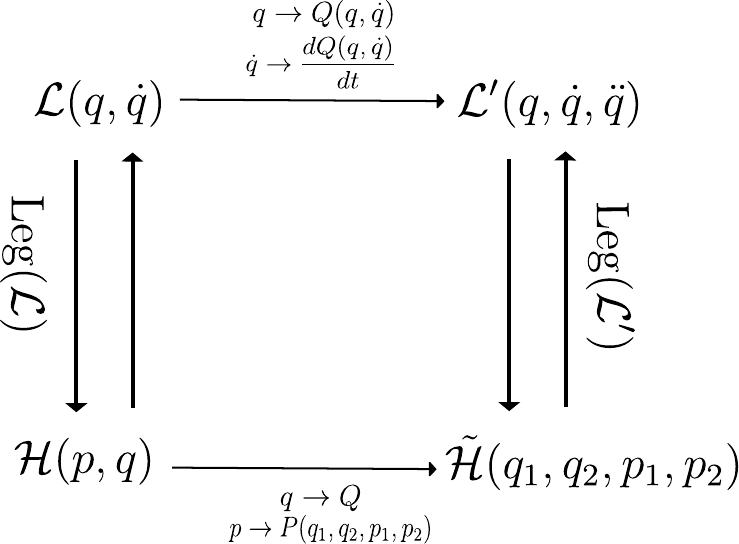}
\caption{\small{Mapping between $S_1$ and $S_2$ in its Lagrangian and Hamiltonian versions.}}
\label{fig:figureII}
\end{center}
\end{figure}

The growth in the order of the action demands the use of extra canonical variables to build the Hamiltonian version of the theory. Throughout the rest of this paper, the Ostrogradskii formalism is applied to construct the Hamiltonian and all the associated conjugate momenta of the new higher-order theory, as seen in \cite{ref:Ostrogradski}. The corresponding Hamiltonian deformation from $\mathcal{H}$ to $\tilde{\mathcal{H}}$ has many peculiar features as will be discussed in the following example. 

Again let us consider the case of a 1D harmonic oscillator as an example:
\begin{equation}
\displaystyle
\mathcal{L}(\phi) = -\frac{1}{2}\phi^T\left(\frac{d^2}{dt^2}+\Omega^2\right)\phi,\label{shoII}
\end{equation}
where the operator $\hat{\mathcal{O}}\chi\equiv \left(d^2/dt^2+\Omega^2\right)\chi$ acts on a test function $\chi$, which is represented by a column vector in the time domain. This operator is separable, which means it can be decomposed as $\hat{\mathcal{O}}=\hat{\mathcal{A}}^T\hat{\mathcal{A}}$ where $\hat{\mathcal{A}}\chi\equiv \left(i\:d/dt+\Omega\right)\chi$ and its transpose $\chi^T\hat{\mathcal{A}}^T\equiv \chi^T\left(i\:d/dt+\Omega\right)^T=\chi^T\left(-i\:d/dt+\Omega\right)$, in the last equality we considered that the time derivative is accurately represented by a skew-symmetric matrix. Moreover, it is possible to notice that the factors of $\hat{\mathcal{O}}$ commute in the following way, $\left[\hat{\mathcal{A}},\hat{\mathcal{A}}^T\right]=0$. A possible way to deform the action is to use the ``square root'' of  $\hat{\mathcal{O}}$ to change the field variable by $\phi\rightarrow \hat{\mathcal{A}}\psi$ and its transpose by $\phi^T\rightarrow \psi^T \hat{\mathcal{A}}^T$. The new Lagrangian now reads
\begin{equation}
\displaystyle
\mathcal{L}'(\psi) = -\frac{1}{2}\psi^T\left[\frac{d^2}{dt^2}+\Omega^2\right]^2\psi,\label{pu}
\end{equation}
which is a particular case of the widely known Pais-Uhlenbeck oscillator as mentioned in \cite{Andrzejewski:2014zia, Chen:2012au}. Strictly speaking, by deforming the field variable we equivalently have deformed $\hat{\mathcal{O}}$ into $\hat{\mathcal{O}}^2$, and by inspecting the trace of the transformation law $\mathrm{tr}\langle \hat{\mathcal{A}}^T\hat{\mathcal{O}}\hat{\mathcal{A}}\rangle = \mathrm{tr}\langle \hat{\mathcal{A}}^T\hat{\mathcal{A}}\hat{\mathcal{O}}\rangle=\mathrm{tr}\langle \hat{\mathcal{O}}^2\rangle$ we notice that a deformation of this type is not unitary, which is of special relevance when the new action has a larger number of degrees of freedom.  After integrating by parts, the Lagrangian in \eqref{pu} can be rewritten as 
\begin{equation}
\displaystyle
\mathcal{L}'(\psi,\dot{\psi},\ddot{\psi}) = -\frac{1}{2}\left[\ddot{\psi}^2-2\Omega^2\dot{\psi}^2+\Omega^4\psi^2\right].\label{pu2}
\end{equation}
The Hamiltonian form of this theory follows from the definition of canonical momenta associated to the relabeled variables $\psi\rightarrow\psi_1$ and $\dot{\psi}\rightarrow\psi_2$
\begin{subequations}
\romansubs
{\allowdisplaybreaks\begin{align}
&\pi_{\psi_1}=\frac{\partial\mathcal{L}'}{\partial\dot{\psi_1}}-\frac{d}{dt}\left(\frac{\partial\mathcal{L}'}{\partial\ddot{\psi_1}}\right)=2\Omega^2\dot{\psi}_1+\Omega^4\dddot{\psi_1},\\
&\pi_{\psi_2}=\frac{\partial\mathcal{L}'}{\partial\dot{\psi_2}}=-\ddot{\psi}_1,
\end{align}}
\end{subequations}
hence the Hamiltonian is given by
\begin{align}
&\mathcal{H}'(\psi_1,\psi_2,\pi_{\psi_1},\pi_{\psi_2}) = \pi_{\psi_1}\psi_2-\frac{1}{2}\pi_{\psi_2}^2\nonumber\\
&-\Omega^2\psi_2^2+\frac{\Omega^4}{2}\psi_1^2.\label{PUhamilton}
\end{align}
Stability issues and ghosts arise immediately: the linear dependence of this expression on $\pi_{\psi_1}$ and its complex conjugate (apart form the negative signs in some of the terms) bring infinitely negative values of energy and unbounded trajectories in phase space. In \cite{Chen:2012au}, unstable solutions are controlled by constraining the number of canonical variables, which can be achieved in both of the limits of the oscillation frequency $\Omega^2$: $\Omega^2\ll 1$ (i.e. suppressing the last term in \eqref{pu2})  and $\Omega^2\gg 1$ (i.e. when the first term in \eqref{pu2} is very small compared with the other two). In both scenarios, there is not much room for an effective reduction of the number of canonical pairs, but this does not have to be the case for dynamical systems with more degrees of freedom. This argument is in agreement with the discussions in \cite{Simon:1990ic} where the stable solutions can always be obtained by varying the model parameters around the limit where two possibilities manifest: (a) there is no contributions from the higher derivative terms or (b) the information carried by the extra derivatives is redundant.  

Another property of the transformations used to generate a higher-order theory can be noticed if we build $\mathcal{L}'(\psi)$ from the original Hamiltonian:
\begin{equation}
\mathcal{H}(\phi,\pi_{\phi})=1/2\left(\pi_{\phi}^2+\Omega^2\phi^2\right) \nonumber
\end{equation}
of the harmonic oscillator by transforming 
\begin{equation}
\pi_\phi\rightarrow P(\psi, \pi_\psi,...)\,, \nonumber 
\end{equation}
and
\begin{equation}
\phi\rightarrow Q(\psi,\pi_\psi) \nonumber
\end{equation}
where the decomposition $P(\psi, \pi_\psi,...)=\pi_\psi+\Delta\pi_\psi$ and $Q(\psi, \pi_\psi,...)=\psi+\Delta\psi$ holds in the same way as in \eqref{defcanonicaH}. By following the lower right corner of Fig.~\ref{fig:figureII}:

\begin{align}
&\Scale[0.90]{P\dot{Q}(\psi,\pi_{\psi})-\frac{1}{2}\left(P^2+\Omega^2Q^2\right)=\mathcal{L}'(\psi_1,\psi_2,\pi_{\psi_1},\pi_{\psi_2})}\nonumber\\
&\Scale[0.90]{\rightarrow P(\psi_1,\psi_2,\pi_{\psi_1},\pi_{\psi_2})=\dot{Q}\pm\sqrt{\dot{Q}^2-2\mathcal{L}'-\Omega^2Q^2}},\nonumber
\end{align}      
which means that the transformation not only brings extra canonical variables but it is also not unique. As noticed in \cite{Woodard:2015zca} these degeneracies are typically used to by-pass Ostrogradskii instabilities. Hence, deformations of this type cannot be confused with coordinate transformations.\\
 
\section{Review of Hamiltonian general relativity and the gauge algebra}\label{sec:algebra}
General relativity is a gauge theory where coordinate freedom is what enables the user to transform results from one coordinate chart to another. As in every gauge theory, it can be equivalently written in the context of a constrained system at the Hamiltonian level in order to define one gauge fixing condition per (first-class) Hamiltonian constraint and find the generators of gauge transformations. As a first step, we will follow the standard way to find the Hamiltonian version of general relativity as discussed in \cite{Poisson:2004tb}, built from the Einstein-Hilbert action
\begin{equation}
S=\int d^4x\sqrt{-g}\:\left[\frac{R}{8\pi G}+\mathcal{L}_m(\psi,g_{\mu\nu})\right],\label{EH}
\end{equation}  
where $\mathcal{L}_m(\psi,g_{\mu\nu})$ is the matter Lagrangian. We will write the spacetime metric using the 3+1 decomposition  
\begin{eqnarray}
&\Scale[0.87]{g_{\mu\nu}=-(N^2-h_{ab}N^aN^b)\delta_\mu^t\delta_\nu^t+2h_{ab}N^b\delta_{(\mu}^t\delta_{\nu)}^a+h_{ab}\delta_\mu^a\delta_\nu^b}\nonumber\\
& \Scale[0.89]{g^{\mu\nu}=-\frac{1}{N^2}\delta^\mu_t\delta^\nu_t+\frac{2N^a}{N^2}\delta^{(\mu}_t\delta^{\nu)}_a+\left(h^{ab}-N^aN^b\right)\delta^\mu_a\delta^\nu_b}\nonumber
\end{eqnarray}
where $N$ and $N^a$ are the lapse function and the shift vector respectively. $h_{ab}$ is the metric of the hypersurface fixed at a constant instant of time. The Gauss-Codazzi equations allow us to write the gravitational part of the action by using the decomposed metric
\begin{equation}
\Scale[0.90]{R=R^{(3)}+K^{ab}K_{ab}-K^2-2\nabla_\alpha\left(n^\beta\nabla_\beta n^\alpha-n^\alpha\nabla_\beta n^\beta\right)}\label{GaussC}
\end{equation} 
where $R^{(3)}$ is the Ricci scalar calculated from $h_{ab}$ and $n^\alpha$ are the components of the normal of the hypersurface at a fixed instant of time and $K^{ab}$ is the extrinsic curvature of the same surface, which is defined as the change of the normal projected by a basis of vectors tangent to the surface. In \eqref{GaussC} the last term between parentheses is a surface term that generally requires cancellation via the addition of the Gibbons-Hawking term. The conjugate momentum to $h_{ab}$ is determined by writing the extrinsic curvature as a function of $\dot{h}_{ab}\equiv \pounds_t h_{ab}$ 
\begin{equation}
K_{ab} = \frac{1}{2N}\left(\dot{h}_{ab}-\nabla_bN_a-\nabla_aN_b\right)\label{extrinsic}  
\end{equation}   
where the metric is Lie transported along a timelike trajectory whose tangent vector is denoted by $t^\alpha$, which is not necessarily parallel to $n^\alpha$, the unit normal of the $t=const.$ surfaces. As in standard field theory, the momentum $\pi_{ab}$ is determined by
\begin{align}
&\pi_{ab}=\frac{\partial}{\partial \dot{h}_{ab}}\left(\frac{\sqrt{-g}R}{8\pi G}\right)= \frac{1}{8\pi G}\frac{\partial K_{cd}}{\partial \dot{h}_{ab}}\frac{\partial}{\partial K_{cd}}\left(\sqrt{-g}R\right)\nonumber\\
&= \frac{\sqrt{h}}{16\pi G}\left(K_{ab}-Kh_{ab}\right).\label{momenta}
\end{align}
Once that the momentum is defined, we can use Legendre transformations to build the total Hamiltonian of the system
\begin{align}
&\Scale[0.90]{\mathcal{H}_T=\displaystyle{\int d^3x \left[\pi^{ab}\dot{h}_{ab}+\pi_{\psi}\dot{\psi} -\left(\frac{\sqrt{-g}R}{8\pi G} + \mathcal{L}_m(\psi, g_{\mu\nu})\right)\right]}}\nonumber\\
&=\displaystyle{\int d^3x \left(N\mathcal{H}_0(x)+ N^a\mathcal{H}_a(x)\right)}\label{scalarvectconst}
\end{align}
($\psi$ representing possible minimally-coupled matter contributions).

The Hamiltonian has been written with respect to $\mathcal{H}_0$ and $\mathcal{H}_a$, which are known as the scalar and vector constraints respectively:
\begin{align}
&\mathcal{H}_0(x) = -\frac{\sqrt{h}R^{(3)}}{16\pi G} +\frac{16\pi G}{\sqrt{h}}\left(\pi^{ab}\pi_{ab}-\frac{1}{2}\pi^2\right) +\mathcal{H}^\psi_0(x)\nonumber\\[0.1cm]
&\mathcal{H}_a(x) = -32\pi G\sqrt{h}\nabla^b\left(\frac{\pi_{ab}}{\sqrt{h}}\right)+\mathcal{H}^\psi_a(x).\label{vector}
\end{align} 
Here $\mathcal{H}^\psi_0$ and $\mathcal{H}^\psi_a$ are the scalar and vector constraints obtained from a matter field (for example, a scalar field). This procedure summarizes the so-called ADM formalism for general relativity \cite{ref:ADM}. In addition to this, the so-called smeared constraints are also important in our discussion, these are defined by
\begin{subequations}
\romansubs
{\allowdisplaybreaks\begin{align}
&\mathcal{H}(N) \equiv \displaystyle{\int d^3x N\mathcal{H}_0(x)},\label{smearedscalar}\\[0.1cm]
&\mathcal{H}(N^a) \equiv \displaystyle{\int d^3x N^a\mathcal{H}_a(x)}.\label{smearedvector}
\end{align}}
\end{subequations}
We must remark that the shift and lapse play the role of Lagrange multipliers since neither $\dot{N}$ nor $\dot{N}^a$ appear explicitly in the action or in the Hamiltonian. Moreover, it is important to figure out if the absence of these terms is not just a gauge artifact. To do so, we must transform both the lapse and the shift vector following the infinitesimal gauge transformation rules of $g^{\mu\nu}$ along an arbitrary vector field $\varepsilon$:
\begin{equation}
\delta_{\varepsilon} g^{\mu\nu} = \frac{\partial{g^{\mu\nu}}}{\partial x^\alpha}\varepsilon^\alpha-g^{\mu\rho}\frac{\partial \varepsilon^\nu}{\partial x^\rho}-g^{\nu\rho}\frac{\partial \varepsilon^\mu}{\partial x^\rho}.\nonumber
\end{equation} 
It is enough to use $\delta_{\varepsilon} g^{00}$ and $\delta_{\varepsilon} g^{0a}$ to determine $\delta_{\varepsilon} N$ and $\delta_{\varepsilon} N^a$ as in \cite{Pons:1996av}, which are given by 
\begin{subequations}
\romansubs
{\allowdisplaybreaks\begin{align}
&\delta_{\varepsilon}N = \frac{\partial N}{\partial x^\mu}\varepsilon^\mu+N\frac{\partial \varepsilon^0}{\partial x^0}-NN^a\frac{\partial \varepsilon^0}{\partial x^a},\label{var1}\\[0.1cm]
&\delta_{\varepsilon}N^a = \frac{\partial N^a}{\partial x^\mu}\varepsilon^\mu+N^a\frac{\partial \varepsilon^0}{\partial x^0}-\left(N^2h^{ab}+N^aN^b\right)\frac{\partial \varepsilon^0}{\partial x^b}\nonumber\\
&\qquad\quad +\frac{\partial \varepsilon^a}{\partial x^0}-N^b\frac{\partial \varepsilon^a}{\partial x^b},\label{var2}
\end{align}}
\end{subequations}
where $N^\mu=N\delta^\mu_0+N^a\delta^\mu_a$ and the total Hamiltonian is $\mathcal{H}(N^\mu)$. We now need to find a solution for $\varepsilon^\mu$ such that $\partial \delta_{\varepsilon}N^\nu/\partial \dot{N}^\mu = 0$, which means that we do not generate any momenta while doing a gauge transformation, which leads us to 4 equations for $\varepsilon^\mu$: 
\begin{align}
\varepsilon^0+N\frac{\partial\varepsilon^0}{\partial N}=&0,\nonumber\\[0.1cm]
\frac{\partial\varepsilon^0}{\partial N^a}=&0,\nonumber\\[0.1cm]
N^a\frac{\partial\varepsilon^0}{\partial N}+\frac{\partial\varepsilon^a}{\partial N}=&0,\nonumber\\[0.1cm]
\varepsilon^0\delta^a_b+\frac{\partial\varepsilon^a}{\partial N^b}=&0.\nonumber
\end{align}
It is now simple to observe that none of these equations depends explicitly on $h_{ab}$, which will be important at the time we perform deformations of the canonical variables. The general solution for this system is $\varepsilon^0 = \xi^0/N$ and $\varepsilon^a = \xi^a-\xi^0N^a/N$, where $\xi^\mu$ is an arbitrary spacetime dependent vector field. We can equivalently use $\xi^\mu$ to represent the same solutions
\begin{equation}
\xi^0=N\varepsilon^0\:;\:\xi^a = \varepsilon^a+N^a\varepsilon^0,\label{trajectories}
\end{equation}
this inversion now makes $\xi^\mu$ a function of $N^\mu$ and defines a new set of coordinates attached to the constant time hypersurface. Hence, it is safe to perform gauge transformations as long as these do not generate momenta of $N^\mu$. Once we identify these vector fields, gauge transformations along any of these solutions are defined just like the equations of motion, following a procedure described in detail in \cite{Rovelli:1986hp}: First, considering that the vector fields $\xi^\mu$ and $\epsilon^\mu$ can be freely exchanged to describe the same gauge flow, we define the Hamiltonian in a way analogous to \eqref{smearedscalar} and \eqref{smearedvector}
\begin{equation}
\mathcal{H}(\xi^\mu)=\tilde{\mathcal{H}}(\varepsilon^\mu)\equiv \displaystyle{\int d^3y \mathcal{H}_\mu(y)\xi^\mu},\nonumber
\end{equation}  
where the integration occurs with respect to the coordinates of the hypersurface. Any arbitrary function $I(h_{ab},\pi_{ab}, \psi, \pi_\psi)$ of the canonical coordinates by using the Poisson brackets
\begin{equation}
\delta_\varepsilon I = \{I,\tilde{\mathcal{H}}(\varepsilon^\mu)\},\label{gauge}
\end{equation}
where $\xi^\mu$ and $\varepsilon^\mu$ are just as defined by \eqref{trajectories}. The gauge algebra acting on the same test function $I$ reads
\begin{equation}
\left(\delta_\varepsilon\delta_\zeta-\delta_\zeta\delta_\varepsilon\right)I= \delta_{[\varepsilon,\zeta]}I.\label{gaugealg}
\end{equation}
Let us evaluate the first two variations on the left
\begin{equation}
\delta_\varepsilon\delta_\zeta I = \left(\delta_\varepsilon N^\mu\frac{\delta}{\delta N^\mu}+\delta_\varepsilon q\frac{\delta}{\delta q}\right)\left(\delta_\zeta q\frac{\delta I}{\delta q}\right),\nonumber
\end{equation}  
where $\delta_\zeta q(\delta I/\delta q)$ is the shorthand notation of
\begin{equation}
\Scale[0.95]{\delta_\zeta q\frac{\delta I}{\delta q}\equiv\delta_\zeta h_{ab}\frac{\delta I}{\delta h_{ab}}+\delta_\zeta \pi_{ab}\frac{\delta I}{\delta \pi_{ab}}+\delta_\zeta \psi\frac{\delta I}{\delta \psi}+\delta_\zeta \pi_{\psi}\frac{\delta I}{\delta \pi_\psi}.}\label{eq:variation2}
\end{equation}
which means that the variation affects all the phase space variables of the system. Using \eqref{gaugealg}, the two variations can be written as 
\begin{equation}
\Scale[0.90]{\delta_\varepsilon\delta_\zeta I = \delta_\varepsilon N^\mu\frac{\delta}{\delta N^\mu}\{I, \tilde{\mathcal{H}}(\zeta)\}+\{\{I,\tilde{\mathcal{H}}(\varepsilon)\},\tilde{\mathcal{H}}(\zeta)\}},\label{lhs}
\end{equation}  
where a substantial difference with respect of other gauge theories comes from the fact that the first term in the right hand side does not cancel. This term is now expressed in detail:
\begin{align}
&\Scale[0.95]{\displaystyle{\delta_\varepsilon N^\mu\frac{\delta}{\delta N^\mu}\{I, \tilde{\mathcal{H}}(\zeta)\} = \delta_\varepsilon N^\mu\frac{\delta}{\delta N^\mu}\int d^3 y z^\alpha\{I, \mathcal{H}_\alpha\}}}\nonumber\\
&\Scale[0.95]{\displaystyle{=\int d^3y \zeta^0\delta_\varepsilon N^\mu\delta^\alpha_\mu\{I, \mathcal{H}_\alpha\}=\{I, \mathcal{\tilde{H}}(\delta_\varepsilon N^\mu\zeta^0)\}}},\label{previous}
\end{align}
where the vector flow $z^\mu:= N\zeta^0\delta^\mu_0+\left(\zeta^a+N^a\zeta^0\right)\delta^\mu_a$ follows from the definition in \eqref{trajectories}. In the last line it is possible to observe that $\partial z^\alpha/\partial N^\mu = \delta^\alpha_\mu\zeta^0$. Therefore the initial variation is given by
\begin{equation}
\delta_\varepsilon\delta_\zeta I = \{\{I,\tilde{\mathcal{H}}(\varepsilon)\},\tilde{\mathcal{H}}(\zeta)\}+\{I, \mathcal{\tilde{H}}(\delta_\varepsilon N^\mu\zeta^0)\}.\nonumber
\end{equation}
With this expression it is possible to rewrite \eqref{gaugealg} as
\begin{align}
&\{\{I,\tilde{\mathcal{H}}(\varepsilon)\},\tilde{\mathcal{H}}(\zeta)\}-\{\{I,\tilde{\mathcal{H}}(\zeta)\},\tilde{\mathcal{H}}(\varepsilon)\}\nonumber\\
&+\{I, \mathcal{\tilde{H}}(\delta_\varepsilon N^\mu\zeta^0-\delta_\zeta M^\mu\varepsilon^0)\}=\{I, \mathcal{\tilde{H}}[\varepsilon,\zeta]\}.\nonumber
\end{align} 
After using the Jacobi identity on the left hand side of the equality, the gauge algebra now reads
\begin{equation}
\{\tilde{\mathcal{H}}(\varepsilon),\tilde{\mathcal{H}}(\zeta)\}=\tilde{\mathcal{H}}([\varepsilon,\zeta]-\delta_\varepsilon N^\mu\zeta^0+\delta_\zeta M^\mu\varepsilon^0).\nonumber
\end{equation}
The other basis of vectors can be used equivalently
\begin{equation}
\Scale[0.95]{\{\mathcal{H}(\xi),\mathcal{H}(z)\}=\mathcal{H}([\xi,z]-\delta_\xi N^\mu z^0+\delta_z M^\mu\xi^0)}\label{constraintalg}.
\end{equation}
In the case of $\xi^\mu=M\delta^\mu_0$ and $z^\mu=N\delta^\mu_0$, the Lie bracket cancels and the first Poisson bracket is given by
\begin{equation}
\{\mathcal{H}(M),\mathcal{H}(N)\}=\mathcal{H}(N\nabla^a M-M\nabla^a N)\label{constraintalg1},
\end{equation}
and in a similar way, the remaining brackets can be determined by choosing $\xi^\mu=M^a\delta^\mu_a$, $z^\mu=N^a\delta^\mu_a$ and $\xi^\mu=M\delta^\mu_0$, $z^\mu=N^a\delta^\mu_a$:
\begin{subequations}
 \romansubs
{\allowdisplaybreaks\begin{align}
&\{\mathcal{H}(M^a),\mathcal{H}(N^a)\}=\mathcal{H}(\pounds_{N^a}M^a)\label{constraintalg2},\\
&\{\mathcal{H}(M),\mathcal{H}(N^a)\}=-\mathcal{H}(\pounds_{N^a}M)\label{constraintalg3}.
\end{align}}
\end{subequations}

The expressions (\ref{constraintalg1}-{}-\ref{constraintalg3}) constitute the constraint algebra of general relativity, which explains the way spacetime contorts as described in \cite{Teitelboim:1972vw}. Strictly speaking, this has spacetime dependent structure constants, so it is not an algebraic structure in the rigorous meaning of the word. An important property of this algebra is its closure, which has been used as a motivation to search for a valid ultraviolet limit of the theory. Nevertheless, our perspective is more conservative and is closely related with the possibility of safely fixing its gauge degrees of freedom. To do so, any choice of gauge should satisfy 
\begin{equation}
\frac{\delta S}{\delta N^\mu}=0 \rightarrow \mathcal{H}_\mu(y)\approx 0,\nonumber
\end{equation}     
which separately implies $\mathcal{H}(N) = \mathcal{H}(N^a) \approx 0$, where ``$\approx$'' means that this statement holds along with the equations of motion. These are also known as on-shell conditions which represent the constrained hypersurfaces where we can find all the possible gauge selections. In order to fix the gauge properly, we must ensure that these surfaces do not evolve in time:
\begin{subequations}
\romansubs
{\allowdisplaybreaks\begin{align}
&\Scale[0.95]{\dot{\mathcal{H}}(M) = \{\mathcal{H}(N),\mathcal{H}(M)\}+\{\mathcal{H}(N^a),\mathcal{H}(M)\}\approx 0},\label{cons1}\\[0.1cm]         
&\Scale[0.95]{\dot{\mathcal{H}}(M^b) = \{\mathcal{H}(N),\mathcal{H}(M^b)\}+\{\mathcal{H}(N^a),\mathcal{H}(M^b)\}\approx 0},\label{cons2}
\end{align}}
\end{subequations}
which are also known as secondary constraints. The closure of the algebra in (\ref{constraintalg1}-{}-\ref{constraintalg3}) ensures that each of the Poisson brackets will always be proportional to other constraints that vanish when evaluated on-shell. In Fig.~\ref{fig:figure3}, we depict the evolution of the hypersurfaces that contain all the possible gauge choices, and the change of any specific choice at a fixed instant of time. Even though a fully detailed discussion on the proper way to do gauge fixing is beyond the scope of our paper, it is important to remark that the gauge degrees of freedom cannot be fixed without this condition. As an additional comment, we must observe that the procedure we followed to derive the algebra is already invariant under coordinate transformations. Later, it will become apparent that one way to keep the same structure under deformations is to introduce a gauge generator per extra degree of freedom introduced. 

\begin{figure}[!h]
\vspace{.5cm}
\begin{center}
\includegraphics[width=0.24\textwidth]{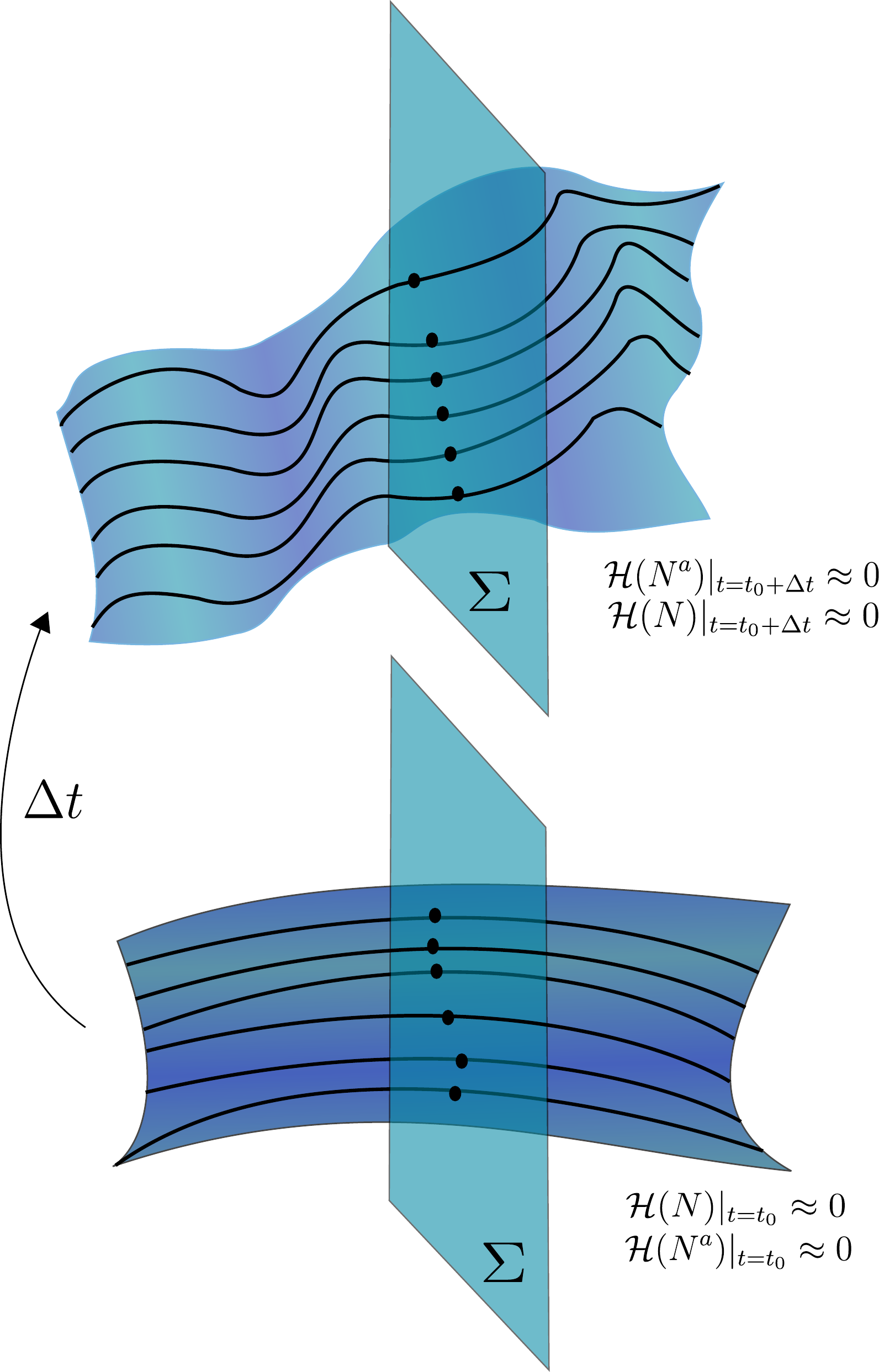}
\caption{\small{Gauge fixing by the intersection with the surface $\Sigma$ as the scalar and vector constraint evolve. The gauge choice changes as the surface (and the orbits) deforms.}}
\label{fig:figure3}
\end{center}
\end{figure}

We need to consider that this procedure is valid at the infinitesimal level. Consequently, as noticed in \cite{Rovelli:1986hp}, one cannot apply the gauge transformations mentioned here to any phase space configuration, since in general these will combine canonical variables (and their derivatives) at different instants of time. Hence, the gauge transformations applied in here can only map a set of solutions of the equations of motion into another version of the same set of solutions. In the subsequent sections, we will translate this procedure to actions that were modified from deforming their canonical variables. It is interesting to explore what kind of transformations of the gauge trajectories in \eqref{trajectories} will transform \eqref{gaugealg} covariantly. The transformed version of \eqref{gaugealg} reads
\begin{equation}
\left(\delta_{\tilde{\varepsilon}}\delta_{\tilde{\zeta}}-\delta_{\tilde{\zeta}}\delta_{\tilde{\varepsilon}}\right)I= \delta_{[\tilde{\varepsilon},\tilde{\zeta}]}I.\nonumber
\end{equation} 
It will be enough to write the first two variations to understand the dependencies of a generic transformation represented in a matrix form by $\tilde{\zeta}= \left(T_{\zeta\rightarrow\tilde{\zeta}}\right)\zeta$ 
\begin{align}
\delta_{\tilde{\varepsilon}}\delta_{\tilde{\zeta}} I = &\left(\left(T_{\varepsilon\rightarrow\tilde{\varepsilon}}\right)\delta_\varepsilon N^\mu\frac{\delta}{\delta N^\mu}+\left(T_{\varepsilon\rightarrow\tilde{\varepsilon}}\right)\delta_\varepsilon q\frac{\delta}{\delta q}\right)\nonumber\\
&\times \left(\left(T_{\zeta\rightarrow\tilde{\zeta}}\right)\delta_\zeta q\frac{\delta I}{\delta q}\right),\nonumber
\end{align}
where the shorthand notation for all the canonical variables $q$ still holds. The only way in which this expression transforms covariantly is if the transformations $T_{\tilde{\zeta}\rightarrow\zeta}$ do not depend on any of the canonical variables (including matter) or on the lapse and shift. This complements the independence on $h_{ab}$ was already used to determine \eqref{trajectories}. Therefore, the smeared constraint algebra in (\ref{constraintalg1}-{}-\ref{constraintalg3}) preserves its form under these conditions.  

\section{Deforming canonical general relativity}\label{sec:GR}
In this section, we deform the gravitational canonical variables $h_{ab}$ and $\pi_{ab}$ following the transformations described in \ref{subsec:transI} and \ref{subsec:transII} from the Einstein-Hilbert action in \eqref{EH} written using the Gauss-Codazzi equations. An equivalent way to express this action is built from the Hamiltonian in ADM variables 
\begin{equation}
\Scale[0.95]{S = \displaystyle{\int dt\int d^3x \left[\pi_\psi\dot{\psi}+\pi^{ab}\dot{h}_{ab}-\mathcal{H}(N)-\mathcal{H}(N^a)\right]}.}\label{actII}
\end{equation}  
In all the cases covered in this project, for simplicity, we will not perform any transformation on the canonical variables of matter. Although this could be done, the particular matter field utilized, and the form of its action, is usually motivated by physics other than gravitational field theory. The purpose of the transformations which follow is to find other theories where both diffeomorphism invariance and the gauge structure are preserved.

\subsection{Deforming general relativity while preserving second-order equations of motion and the number of degrees of freedom}\label{subsec:canon}
In this section, we will show that these transformations lead us unavoidably to Lovelock's theorem. Which shows that the only second-order curvature based metric gravitational theory equipped with diffeomorphism invariance is general relativity. In order to do that, we deform the canonical variables $h_{ab}$ and $\pi_{ab}$ in \eqref{EH} by considering a transformation of variables analog to what was presented in \ref{subsec:transI}, which in the Hamiltonian formalism would be  
\begin{subequations}
\romansubs
{\allowdisplaybreaks\begin{align}
&h_{ab}\rightarrow H_{ab}(\tilde{h}_{ab},\tilde{\pi}_{ab})=\tilde{h}_{ab}+\Delta\tilde{h}_{ab},\label{ct1}\\[0.1cm]
&\pi_{ab}\rightarrow P_{ab}(\tilde{h}_{ab},\tilde{\pi}_{ab})=\tilde{\pi}_{ab}+\Delta\tilde{\pi}_{ab},\label{ct2}
\end{align}}
\end{subequations}
which correspond to the transformations of the variables $h_{ab}$ and $\dot{h}_{ab}$ in the Lagrangian. Notice that in the right hand side of both expressions the terms were expanded in the same way as in \eqref{defcanonicaH}. In its simplest version, we can consider the deviations of these variables as $\Delta\tilde{h}_{ab} = H_{ab}(\tilde{h}_{ab},\tilde{\pi}_{ab}) - \tilde{h}_{ab}$ and $\Delta\tilde{\pi}_{ab} = P_{ab}(\tilde{h}_{ab},\tilde{\pi}_{ab}) - \tilde{\pi}_{ab}$ without any further assumptions on the magnitudes of $\Delta\tilde{h}_{ab}$ and $\Delta\tilde{\pi}_{ab}$. After these deformations the action in \eqref{actII} is given by
\begin{align}
S' = &\displaystyle{\int dt\int d^3x \left[\pi_\psi\dot{\psi}+\tilde{\pi}^{ab}\dot{\tilde{h}}_{ab}-\tilde{\mathcal{H}}(N)-\tilde{\mathcal{H}}(N^a)\right]}\nonumber\\
 &+\Delta\mathcal{L}(\tilde{h}_{ab},\tilde{\pi}_{ab},\Delta\tilde{h}_{ab},\Delta\tilde{\pi}_{ab},\psi).\label{actcan}
\end{align}  
The term $\Delta\mathcal{L}$ contains all of the terms proportional that contain some power of $\Delta\tilde{h}_{ab}$ and $\Delta\tilde{\pi}_{ab}$. We will count degrees of freedom in the same manner as in \cite{Henneaux:1992ig}:
\begin{align}
&2\times\{\#\:\mathrm{of\:degrees\:of\:freedom}\} = \nonumber\\[0.1cm]
&\{\#\:\mathrm{of\:canonical\:variables}\}  \nonumber\\
&-2\times\{\#\:\mathrm{of\:first\:class\:constraints}\} \nonumber\\
& -2\times\{\#\mathrm{of\:second\:class\:constraints}\}.
\end{align}  
GR does not have second class constraints. If the number of canonical variables has not changed and/or if the new theory remains being a metric theory, there is no reason to expect any change in the number of degrees of freedom. Also, we expect that the existing constraints commute on-shell with each other otherwise diffeomorphism invariance is broken. Since we require the presence of that symmetry, none of the constraints should be demoted to second-order.  Now, the procedure suggested by Dirac in \cite{Dirac:1951zz} allows us to decompose $\Delta\mathcal{L}$ in \eqref{actcan} as
\begin{equation}
\Delta\mathcal{L} = N^\mu V_\mu,\label{decompose}
\end{equation}  
this means that it is possible find a vector -{}- in the basis formed by the normal and the triad vectors tangential to the hypersurface -{}- in which the extra piece of the action can be reprojected. It is possible therefore, to rewrite the action in \eqref{actcan} with respect to a new Hamiltonian
\begin{equation}
\Scale[0.90]{S' = \displaystyle{\int dt\int d^3x \left[\pi_\psi\dot{\psi}+\tilde{\pi}^{ab}\dot{\tilde{h}}_{ab}-\hat{\mathcal{H}}(N)-\hat{\mathcal{H}}(N^a)\right]},}\label{actcan2}
\end{equation}  
where the new scalar and vector constraints are given by
\begin{equation}
\hat{\mathcal{H}}(N) =\tilde{\mathcal{H}}(N)+ N V_0\:,\: \hat{\mathcal{H}}(N^a) =\tilde{\mathcal{H}}(N^a)+ N^a V_a.\nonumber
\end{equation}  
If we demand that the new Hamiltonian constraints satisfy the gauge algebra in (\ref{constraintalg1}-{}-\ref{constraintalg3}) and the condition in \eqref{decompose} we have 4 conditions for the four components of $V^\mu$. If there is not a unique solution, the system has more degrees of freedom than the ones already counted, which would be a contradiction. Therefore, we will assume that there is a unique solution for $V^\mu$. 
On the other hand, the Lie derivative of a scalar function $F$ along the time direction $t$ is given by
\begin{equation}
\Scale[0.95]{\pounds_t F = t^\alpha \nabla_\alpha F = (N n^\alpha + N^a e^\alpha_a)\nabla_\alpha F \equiv N^\kappa \nabla_\kappa F,}\label{derext}
\end{equation}  
where we selected a basis in which, by definition, the normal defines an orthogonal coordinate to the surface. This works in the same way as the basis that allows us to write the four indices in $N^\mu$. A term like this would correspond to the time derivative of the generator of canonical transformations. Such a modification would only make the new action different from general relativity by a total derivative and hence produce identical equations of motion. A comparison of \eqref{decompose} with the last expression reveals that it is not possible to generate $V^\mu$ via the ``gradient'' in this basis, then this means that the vector $V^\mu$ needs an extra ``solenoidal'' current to be reconstructed. The existence of an extra current would be a clear indication of extra degrees of freedom, which severely contradicts the counting previously made. So the only possibility we have is that the new action is different from the one in \eqref{EH} just by a derivative along the time direction. This means that if one wants to preserve the symmetry and the number of degrees of freedom after transforming the canonical variables, there is no other option than a canonical transformation. This is fully consistent with Lovelock's theorem, just as stated in \cite{Clifton:2011jh}.

\subsection{Deforming general relativity by introducing extra degrees of freedom}
In this section, we perform a concrete implementation of the transformations presented in \ref{subsec:transII} for the Einstein-Hilbert action. This is arguably the most complex type of deformation and therefore will take up the bulk of the analysis. 

\subsubsection{Introduction}
In the following, the theory of deformation of variables is introduced for gravitation in the usual degrees of freedom. Using Weinberg's
``Folk Theorem'' \cite{1979PhyA...96..327W, Weinberg:2016kyd}, one can construct the most general effective field theory by constructing a Lagrangian that contains all possible
diffeomorphism invariant terms, using only the degrees of freedom of the theory. Let $\mathcal{L}_{\text{D}}$ be the most general diffeomorphism invariant theory made up of the curvature. This would produce the following expansion, 
\begin{equation}
\mathcal{L}_{\text{D}}=\alpha R+\left(\beta_{1}R^{2}+\beta_{2}R_{ab}R^{ab}+\cdots\right)+\mathcal{O}\left(R^{3}\right)\label{eq:mike-folktheorem}
\end{equation}

where the first term would correspond to GR. The objective is to explore another sector of diffeomorphism invariant theories by means of the method of variable deformation, in which one makes a replacement of the coordinate $h_{ab}$ by some function of the coordinates $H_{ab}$,
such as below
\begin{equation}
h_{ab}\rightarrow H_{ab}\left(h_{cd},\;\lie_{t}h_{cd}\right)\label{eq:mike-deformationforthistheory}
\end{equation}

where $H_{ab}$ is a mapping that can not be a canonical transformation of the original ADM variables. In total generality, $H_{ab}$ will
acquire its space-time dependence through the degrees of freedom ($h_{ab}(x)$, $\lie_{t}h_{ab}(x)$, etc.),
but can also have its own spacetime dependence apart from this, which will be addressed via an example in the next subsections.
Due to the additional time derivatives found in terms like the intrinsic curvature $K_{ab}$, this will produce a theory dependent on higher
order derivatives than what is found in GR. This feature appears in $R^{2}$ gravity as well \cite{Grav1987}. In general, this deformation can
depend on even higher order time derivatives of the 3-metric,
\begin{equation}
h_{ab}\rightarrow H_{ab}\left(h_{cd},\lie_{t}h_{cd},\lie_{t}\lie_{t}h_{cd},\cdots\right)\label{eq:Mike-general deformation}
\end{equation}

which will provide a theory with higher in order of time derivatives. The volume element will not be deformed due to geometric reasons; The measure of all parts of the action must remain the infinitesimal volume element for it to remain a proper action. These theories provide an expanded sector of diffeomorphism invariant theories,
and is depicted in Figure \ref{fig:Depiction_of_deformation}. 

\begin{figure}[!ht]
\vspace{.5cm}
\begin{center}
\includegraphics[width=0.50\textwidth]{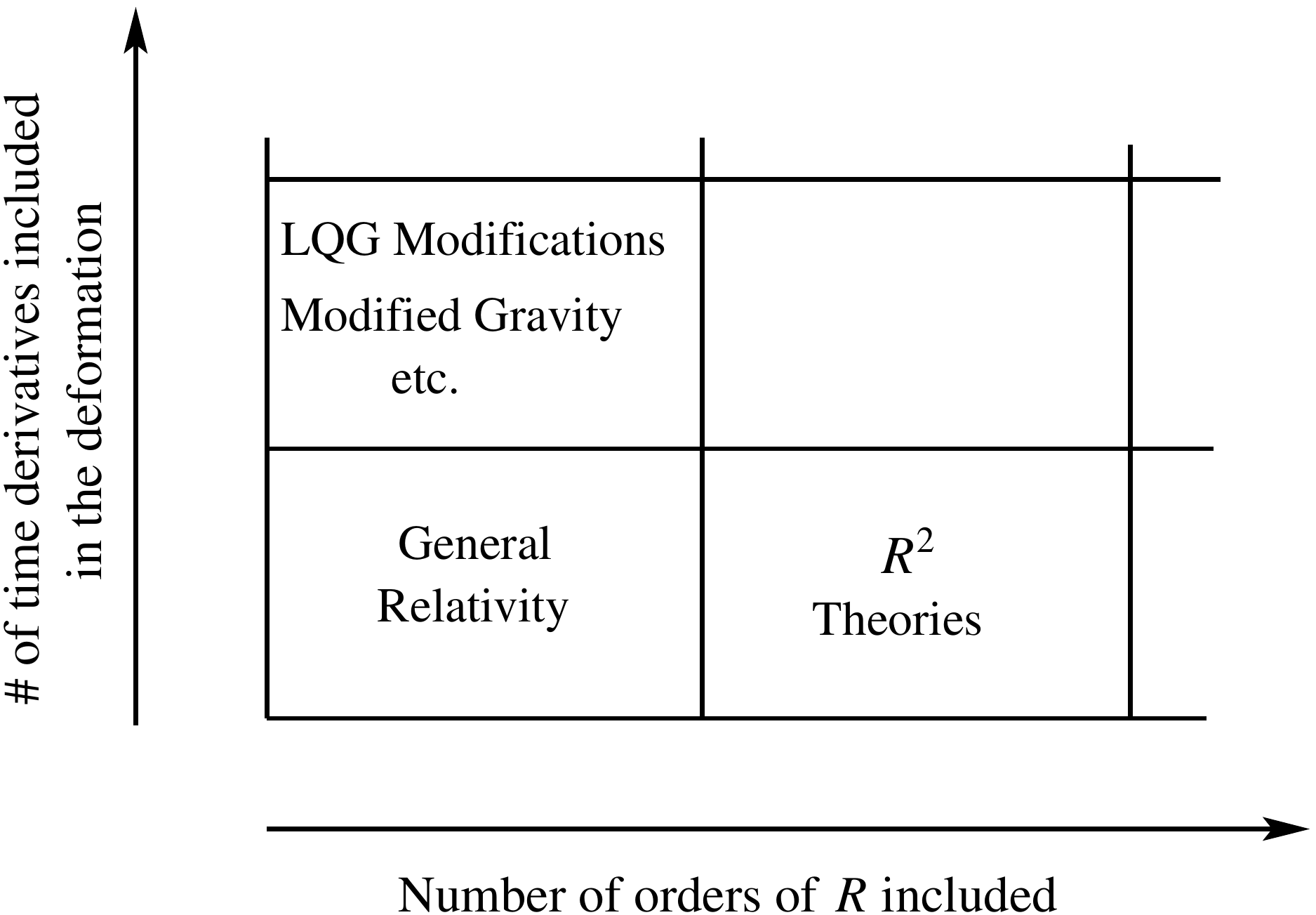}
\caption{{\small{Depiction of deformation of variables. The rows determine the number of terms included from equation $\eqref{eq:mike-folktheorem}$, while the columns determine how many variables you include in the deformation $h_{ab}\rightarrow H_{ab}\left(h_{cd},\protect\lie_{t}h_{cd}, \protect\lie_{t}\protect\lie_{t}h_{cd},\cdots\right)$. The focus of the rest of this manuscript will deal with the top left sector.}}}
\label{fig:Depiction_of_deformation}
\end{center}
\end{figure}

To gain intuition for the effects of the deformation, this work will consider only the Einstein-Hilbert term of the expansion $\eqref{eq:mike-folktheorem}$, with the deformation $\eqref{eq:mike-deformationforthistheory}$ as this is arguably the most interesting case.

\subsubsection{The new degrees of freedom}

After deforming via a substitution like in equation $\eqref{eq:Mike-general deformation}$, the theory will generally change. We wish to maintain the 3-metric, $h_{ab}$, as the configuration variable which dictates the geometric features of spacetime. It is for this reason that the volume element $\sqrt{-g}$ will not be deformed via  $\eqref{eq:Mike-general deformation}$ or by any other means. Any other variations due to the deformation $H_{ab}$ will serve to change the equations
of motion for $h_{ab}$, which as stated is to be maintained as the configuration variable, through its alteration of the theory. (That is, we deform a metric theory into another metric theory.) Another point is that since $H_{ab}$ depends on time derivatives of $h_{ab}$, this theory will now depend on accelerations, jerks, etc, instead of the usual dependence on the field and its velocity field. This
introduces a new degree of freedom for every step we take vertically in Figure $\ref{fig:Depiction_of_deformation}$ and must be addressed. 

\subsubsection{Gauge Invariance and Ostrogradskiian Instabilities}

The essence of GR is arguably its gauge invariance, which, like any other gauge theory, generates the structure of the interactions in
the theory. In gravity, the gauge transformation that must be invariant is that of the diffeomorphism, which endows the theory with the property that leaves the mechanics of gravity the same regardless of any specific location in spacetime (known as background independence). To verify this gauge invariance, one should obtain the constraints of this new theory, and what further constraints need to be satisfied. It shall be shown below, for a deformed theory, that in general there will be additional constraints generated on top of the usual scalar and vector constraints. 

One further point that requires clarification is that of the Dirac's ``Constraint Algebra''\cite{Dirac:1951zz} for which the scalar and vector constraints are elements, will be satisfied regardless of what their form is. This algebraic structure was shown by Rovelli to be directly derived from the diffeomorphism gauge algebra \cite{Rovelli:1986hp}. What this means is that whether one is dealing with the scalar and vector constraints for a deformed theory or a higher order theory like the $R^{2}$ variety, the ``constraint algebra'' should be automatically verified. 

Obtaining these constraints, namely the scalar, vector and ``additional'' constraints, requires that the theory is phrased in Hamiltonian form, which in our case depends not only on the velocity and position field coordinates, but perhaps also that of the acceleration, jerk, etc. Typically, such theories are energetically not bound from below and thus can infinitely go into energetic debt by borrowing more and more energy to create unstable results. Ostrogradskii provided a treatment of this situation for theories without gauge ambiguities, but since gravity is a gauge theory, a modification of this treatment is required and was provided by Buchbinder, Gitman, Lyakovich and Tyutin \cite{I-1985, D-1983} and applied to $R^{2}$ gravity \cite{Grav1987}. The summary of this approach is that one can define a new canonical variable such that its time derivatives can ``absorb'' any higher-derivative term -{}- and potential instabilities -{}-.This procedure can be generalized as in \cite{D-1983}, and the unstable nature of Ostrogradskiian theories will also be addressed in what follows. For the theory proposed, which is the Einstein-Hilbert action deformed via $\eqref{eq:mike-deformationforthistheory}$, one can decompose
the deformation in the following way,
\begin{eqnarray}
&H_{ab}\left(h_{cd},\lie_{t}h_{cd}\right)=h_{ab}+\tilde{H}_{ab}\left(h_{cd}\right)+\nonumber\\
&J_{ab}\left(h_{cd},\lie_{t}h_{cd}\right)\label{eq:mike-deformation decomposition}
\end{eqnarray}

where $\tilde{H}_{ab}\left(h_{cd}\right)$ is the part of the general deformation that conformally transforms $h_{ab}$ in a spacetime
dependent way. This can be interpreted as the gravitational analogue to the deformation that lead us to \eqref{pu}. Moreover, the deformations can also be used to represent semiclassical contributions. Later in this paper, we will extend these procedures to the Ashtekar/Tetrad formalism in section \ref{sec:Ashtekar}. In the latter expression, the next term, $J_{ab}$ depends strictly on higher order derivatives and is the natural choice for a new canonical variable as its time derivative depends on higher order time derivatives of $h_{ab}$
\begin{equation}
\dot{J}_{ab}=\frac{\partial J_{ab}}{\partial h_{cd}}\dot{h}_{cd}+\frac{\partial J_{ab}}{\partial\dot{h}_{cd}}\ddot{h}_{cd} \nonumber
\end{equation}
which absorbs the higher order time derivatives found in the new gravitation Lagrangian. It should be recalled that only $h_{ab}$
is used to contract vectors in our spacetime, while the other terms serve to deform the theory. At this point it is crucial to note that
the relation between $J_{ab}$ and the usual degrees of freedom, which is encoded in $\frac{\partial J_{ab}}{\partial h_{cd}}$ and $\frac{\partial J_{ab}}{\partial\dot{h}_{cd}}$ is determined by diffeomorphism invariance. Thus, the deformation $H_{ab}$ must satisfy a certain functional form to be diffeomorphism invariant. This fact will be demonstrated in the next subsections.

\subsubsection{The form of the deformed theory}

After the deformation $\eqref{eq:mike-deformationforthistheory}$,
the Lagrangian will look as follows:
\begin{eqnarray*}
& \frac{16\pi}{\sqrt{-g}}\mathcal{L} =R =R^{\left(3\right)}+K^{ac}K_{ac}-K^{2}\\
&-2\nabla_{\alpha}\left(\tudu n{\alpha}{;\beta}{}{}n^{\beta}-\tudu n{\beta}{;\alpha}{}{}n^{\alpha}\right) +16\pi\mathcal{L}_{\text{MG}}\left(h,\dot{h},\ddot{h}\right)
\end{eqnarray*}
where everything written is the standard GR Lagrangian plus $\mathcal{L}_{\text{MG}}$
which represents the modifications due to the deformation,
\begin{eqnarray*}
&16\pi\mathcal{L}_{\text{MG}}\left(h,\dot{h},\ddot{h}\right) =R_{\tilde{H}}^{\left(3\right)}+R_{J}^{\left(3\right)}+V_{H}\left(\tilde{H},h\right)\\
&+V_{J}\left(J,h\right)+ \dot{J}\left(\frac{K}{N}+\frac{1}{4N^{2}}\left(2\frac{\partial\tdud{\tilde{H}}{}{}{}{}}{\partial h_{ef}}\dot{h}_{ef}\right)\right)+\frac{1}{4N^{2}}\dot{J}^{2}+\\
& \frac{1}{2N^{2}}\left(2NK^{ab}+\frac{\partial\tilde{H}^{ab}}{\partial h_{ef}}\dot{h}_{ef}\right)\dot{J}_{ab}+\frac{1}{4N^{2}}h^{ac}h^{bd}\dot{J}_{ab}\dot{J}_{cd}+\\
& \dot{h}_{cd}\left(\frac{K}{N}\frac{\partial\tdud{\tilde{H}}{}{}{}{}}{\partial h_{cd}}+\frac{1}{N}\frac{\partial\tilde{H}_{ab}}{\partial h_{cd}}K^{ab}\right)+\frac{1}{4N^{2}}\bigg(\frac{\partial\tdud{\tilde{H}}{}{}{}{}}{\partial h_{cd}}\frac{\partial\tdud{\tilde{H}}{}{}{}{}}{\partial h_{ef}}+\\
&\frac{\partial\tilde{H}^{cd}}{\partial h_{gh}}\frac{\partial\tilde{H}_{cd}}{\partial h_{ef}}\bigg)\dot{h}_{ef}\dot{h}_{cd}
\end{eqnarray*}
where the first terms come from inserting the decomposition $\eqref{eq:mike-deformation decomposition}$
in the definition for $R$, which gives
\[
\tilde{R}^{\left(3\right)}=R^{\left(3\right)}+R_{\tilde{H}}^{\left(3\right)}+R_{J}^{\left(3\right)}+V_{H}\left(\tilde{H},h\right)+V_{J}\left(J,h\right)
\]

In order to map the Lagrangian theory to the Hamiltonian form, the conjugate momenta to $h_{ab}$ and $J_{ab}$ must be found so that
the Legendre transformation is defined. They are, 

\begin{align}
	\pi^{ab} & =\frac{1}{16\pi}\frac{\partial\left(16\pi\sqrt{-g}\mathcal{L}_{G}\right)}{\partial\dot{h}_{ab}}\nonumber \\
	& =\underbrace{\frac{1}{16\pi}\sqrt{h}\left(K^{cd}-h^{cd}K\right)}_{\pi_{0}^{ab}}\nonumber \\
	+ & \frac{\sqrt{h}}{32\pi N}\frac{\partial\dot{J}_{cd}}{\partial\dot{h}_{ab}}\left(2NKh^{cd}+2NK^{cd}\right)\nonumber \\
	+ & \frac{\sqrt{h}}{32\pi N}\dot{h}_{cd}\left(h^{ab}\frac{\partial\tdud{\tilde{H}}{}{}{}{}}{\partial h_{cd}}+\frac{\partial\tilde{H}^{ab}}{\partial h_{cd}}+\frac{\partial\tdud{\tilde{H}}{}{}{}{}}{\partial h_{cd}}\frac{\partial\tdud{\tilde{H}}{}{}{}{}}{\partial h_{ab}}+\frac{\partial\tilde{H}^{\tilde{c}\tilde{d}}}{\partial h_{cd}}\frac{\partial\tilde{H}_{\tilde{c}\tilde{d}}}{\partial h_{ab}}\right)\nonumber\\
	+ & \frac{\sqrt{h}}{32\pi N}\frac{\partial\dot{J}_{cd}}{\partial\dot{h}_{ab}}\left(h^{cd}\frac{\partial\tdud{\tilde{H}}{}{}{}{}}{\partial h_{ef}}\dot{h}_{ef}+\frac{\partial\tilde{H}^{cd}}{\partial h_{ef}}\dot{h}_{ef}\right)\nonumber\\
	+ & \frac{\sqrt{h}}{32\pi N}\left(2NK\frac{\partial\tdud{\tilde{H}}{}{}{}{}}{\partial h_{ab}}+2NK^{cd}\frac{\partial\tilde{H}_{cd}}{\partial h_{ab}}\right)\nonumber\\
	+ &\frac{\sqrt{h}}{32\pi N}\dot{J}\left(h^{ab}+\left(\frac{\partial\tdud{\tilde{H}}{}{}{}{}}{\partial h_{ab}}\right)+\frac{\partial\dot{J}}{\partial\dot{h}_{ab}}\right)\nonumber\\
	+&\frac{\sqrt{h}}{32\pi N}\left(\dot{J}_{ab}\frac{\partial\tilde{H}^{cd}}{\partial h_{ab}}\dot{J}_{cd}+\frac{\partial\dot{J}_{cd}}{\partial\dot{h}_{ab}}\dot{J}^{cd}\right)\label{eq:mike-pih}
\end{align}
\begin{align}
	\pi_{J}^{ab} & =\frac{1}{16\pi}\frac{\partial\left(16\pi\sqrt{-g}\mathcal{L}_{G}\right)}{\partial\dot{J}_{ab}}\nonumber \\
	& =\frac{\sqrt{h}}{32\pi N}\left(h^{ab}\frac{\partial\tdud{\tilde{H}}{}{}{}{}}{\partial h_{ef}}\dot{h}_{ef}+\frac{\partial\tilde{H}^{ab}}{\partial h_{ef}}\dot{h}_{ef}\right)\nonumber\\
	+ & \frac{\sqrt{h}}{32\pi N}\left(2NK^{ab}+h^{ab}2NK\right)+\nonumber\\
	+ & \frac{\sqrt{h}}{32\pi N}\left(\dot{J}^{ab}+h^{ab}\dot{J}\right)\label{eq:mike-piJ} 
\end{align}

Let $\mathcal{L}_{\text{G}}^{0}$ be the original Lagrangian for GR, then the new Hamiltonian density is,
\begin{align}
	\mathscr{H}_{\text{MG}} & =\pi^{ab}\dot{h}_{ab}+\pi_{J}^{ab}\dot{J}_{ab}-N\sqrt{h}\mathcal{L}_{G}\nonumber \\
	& =-N\sqrt{h}\mathcal{L}_{G}^{0}\nonumber \\
	- & \frac{\sqrt{h}N}{16\pi}\left(R_{\tilde{H}}^{\left(3\right)}+V_{H}\left(\tilde{H},h\right)+V_{J}\left(J,h\right)\right)\nonumber \\
	+ & \frac{\sqrt{h}}{32\pi N}\left(R_{J}^{\left(3\right)}+\frac{1}{2}\dot{J}^{ab}\dot{J}_{ab}+\frac{1}{2}\dot{J}^{2}\right)\nonumber \\
	+ & \frac{\sqrt{h}}{32\pi N}\bigg(2NKh^{cd}+2NK^{cd}+h^{cd}\frac{\partial\tdud{\tilde{H}}{}{}{}{}}{\partial h_{ef}}\dot{h}_{ef}\nonumber\\
	+ & \frac{\partial\tilde{H}^{cd}}{\partial h_{ef}}\dot{h}_{ef}\bigg)\frac{\partial\dot{J}_{cd}}{\partial\dot{h}_{ab}}\dot{h}_{ab}\nonumber\\
	+ & \frac{\sqrt{h}}{32\pi N}\bigg(\frac{\partial\tilde{H}^{cd}}{\partial h_{ab}}\dot{h}_{ab}\dot{J}_{cd}+\frac{\partial\tdud{\tilde{H}}{}{}{}{}}{\partial h_{ab}}\dot{h}_{ab}\dot{J}+\dot{J}\frac{\partial\dot{J}}{\partial\dot{h}_{ab}}\dot{h}_{ab}\nonumber\\
	+ & \frac{\partial\dot{J}_{cd}}{\partial\dot{h}_{ab}}\dot{h}_{ab}\dot{J}^{cd}\bigg)\nonumber \\
	+ & \frac{\sqrt{h}}{32\pi N}\left(\dot{J}^{ab}+h^{ab}\dot{J}\right)\dot{h}_{ab}\nonumber \\
	+ & \frac{\sqrt{h}}{32\pi N}\dot{h}_{ab}\dot{h}_{cd}\bigg(h^{ab}\frac{\partial\tdud{\tilde{H}}{}{}{}{}}{\partial h_{cd}}+\frac{\partial\tilde{H}^{ab}}{\partial h_{cd}}+\frac{1}{2}\frac{\partial\tdud{\tilde{H}}{}{}{}{}}{\partial h_{cd}}\frac{\partial\tdud{\tilde{H}}{}{}{}{}}{\partial h_{ab}}\nonumber\\
	+ & \frac{1}{2}\frac{\partial\tilde{H}^{\tilde{c}\tilde{d}}}{\partial h_{cd}}\frac{\partial\tilde{H}_{\tilde{c}\tilde{d}}}{\partial h_{ab}}\bigg)\label{eq:Mike-Hmg} 
\end{align}

In the usual Hamiltonian treatment, the given theory is written strictly in terms of the canonical field variables $\phi_{\alpha}$ and $\pi^{\alpha}=\frac{\partial\mathcal{L}}{\partial\dot{\phi}_{\alpha}}$ \cite{Dirac:1951zz}, which requires, at minimum, a surjective mapping from the velocities to the canonical momentum. In the general case depicted in equation $\eqref{eq:Mike-Hmg}$, one would need both the squares of the two new momenta $\pi_{ab}\pi^{ab}\thinspace\thinspace,\thinspace\thinspace\pi_{J}^{ab}\pi_{J,ab}$ and the squares of the traces $\pi^{2}$and $\pi_{J}^{2}$ so that one can write this as 
\begin{eqnarray*}
&\mathscr{H}_{\text{MG}}=\frac{N}{\sqrt{h}}\left(\pi_{J}^{ab}\pi_{J,ab}+\pi^{ab}\pi_{ab}\right)+\\
&\mathscr{H}_{\text{MG,2}}\left(J,\dot{J},h,\dot{h},\pi,\pi_{J}\right)
\end{eqnarray*}
where whether the mapping between these two momenta and the two velocities $\dot{h}_{ab}$ and $\dot{J}_{ab}$ is injective or not will determine whether one can write $\mathcal{\mathscr{H}_{\text{MG,2}}}$ strictly in terms of the new conjugate variables. The computation for the general Hamiltonian density in terms of the momenta but not the velocities is long and only truly required if we want the equations of motion for this theory, which we do not at this moment.

\subsubsection{An example}

Equation $\eqref{eq:Mike-Hmg}$ has clear kinetic energy terms for different tensorial components of $J_{ab}$, which is seen in the $\left(R_{J}^{\left(3\right)}+\frac{1}{2}\dot{J}^{ab}\dot{J}_{ab}+\frac{1}{2}\dot{J}^{2}\right)$ term. The form of the Hamiltonian is similar to that of a scalar field, i.e. , the $\dot{J}^{2}$ term, and a tensor field, through $\dot{J}_{ab}\dot{J}^{ab}$. These can be interpreted as different sectors of the gravitational field that come as a result of this deformation. Moreover, these deformations have less degrees of freedom than the original gravitational variables in order to circumvent the potential instabilities created by introducing higher-derivatives, as discussed in \cite{Woodard:2015zca}. 

Instead of acquiring the constraints in general, it is instructive to decompose $J_{ab}$ and $\dot{J}_{ab}$ into scalar, vector and tensor
components, 
\[
J_{ab}=Jh_{ab}+J_{ab}^{\text{V}}+J_{ab}^{\text{T}}
\]
\[
\dot{J}_{ab}=\dot{J}h_{ab}+\dot{J}_{ab}^{\text{V}}+\dot{J}_{ab}^{\text{T}}
\]
and then consider the dynamics due to only a subset of these degrees of freedom. Example routes are outlined in Table $\eqref{tab:Mike-Jdegreeoffreedom}$.

\begin{table}[!ht]
\[
\begin{array}{r|l|l}
\underline{\text{\small{Degs. of Freedom for} }J_{ab}} & \underline{\dot{J}^{2}\text{ {\small{Value}}}} & \underline{\dot{J}_{ab}\dot{J}^{ab}\text{ {\small{Value}}}}\\
\text{Only Scalar} & \dot{J}^{2}\neq0 & \dot{J}_{ab}\dot{J}^{ab}=h\dot{J}^{2}\\
\text{Only Vector\textbackslash Tensor} & \dot{J}^{2}=0 & \dot{J}_{ab}\dot{J}^{ab}=\\
&  & \Scale[0.92]{\left(J_{\text{tensor}}^{ab}+J_{\text{vector}}^{ab}\right)^{2}}\\
\text{\ensuremath{\underset{\left(Special\right)}{\text{Scalar-Vector-Tensor}}}} & \dot{J}^{2}\neq0 & \dot{J}_{ab}\dot{J}^{ab}=0\\
\underset{\text{\ensuremath{\left(General\right)}}}{\text{Scalar-Vector-Tensor}} & \dot{J}^{2}\neq0 & \dot{J}_{ab}\dot{J}^{ab}\neq0
\end{array}
\]
	
	\caption{\label{tab:Mike-Jdegreeoffreedom}{{\small{The various ways to decompose the
		tensorial structure of the new degree of freedom $J_{ab}$.}}}}
\end{table}

The simplest case is that where the scalar degree of freedom vanishes, i.e., the trace vanishes $\dot{J}=0$, but the tensorial structure is
maintained. It will be assumed that the natural conformal scaling due to $\tilde{H}$ in equation $\eqref{eq:mike-deformation decomposition}$ will be suppressed to zero. In this way, only the new coordinate $J_{ab}$ deforms the theory. 

In this example the conjugate momenta simplify to 
\pagebreak
\begin{strip}
\begin{align}
\pi^{ab} & =\frac{1}{16\pi}\sqrt{h}\left(K^{cd}-h^{cd}K\right) \nonumber\\
& \frac{\sqrt{h}}{16\pi}\frac{\partial\dot{J}_{cd}}{\partial\dot{h}_{ab}}\left[\frac{16\pi}{\sqrt{h}}\pi_{J}^{cd}-\frac{1}{2N}\dot{J}^{cd}\right] \nonumber \\
& \frac{\sqrt{h}}{32\pi N}\left(\dot{J}_{ab}+\frac{\partial\dot{J}_{cd}}{\partial\dot{h}_{ab}}\dot{J}^{cd}\right) \nonumber\\[0.05cm]	
\pi_{J}^{ab} & =+\frac{\sqrt{h}}{16\pi}\left(K^{ab}+h^{ab}K\right)+\frac{\sqrt{h}}{32\pi N}\left(\dot{J}^{ab}\right) \nonumber 
\end{align}


\begin{align}
	\mathcal{H}_{G} & \Scale[0.90]{=\frac{\sqrt{h}}{16\pi}\bigg[\left(\frac{16\pi}{\sqrt{h}}\right)^{2}\left[\pi^{ab}\pi_{ab}-\frac{\tilde{\pi}^{2}}{2}\right]-R^{\left(3\right)} +\left(\frac{16\pi}{\sqrt{h}}\right)^{2}\left[\pi_{J}^{ab}\pi_{J,ab}-\frac{\tilde{\pi}_{J}^{2}}{2}\right]-R_{J}^{\left(3\right)}+\frac{1}{2}\left(\frac{16\pi}{\sqrt{h}}\right)^{2}\tilde{\pi}\tilde{\pi}_{J}-V_{J}\left(J,h\right)\bigg]N} \nonumber\\
	& \Scale[0.97]{+\frac{\sqrt{h}}{16\pi}\left[2K_{1}K_{1}^{ab}\frac{\partial\dot{J}_{ab}}{\partial\dot{h}^{cd}}h^{cd}-\left[2\frac{\partial\dot{J}_{cd}}{\partial\dot{h}^{ab}}+\frac{\partial\dot{J}_{ab}}{\partial\dot{h}^{ef}}\frac{\partial\dot{J}_{cd}}{\partial\dot{h}_{ef}}\right]K_{1}^{cd}K_{1}^{ab}-K_{1}^{ab}K_{1ab}-5K_{1}^{2}\right]N}\nonumber \\
	& \Scale[0.97]{+\frac{\sqrt{h}}{16\pi}\left[-2\left(K_{1}^{ab}-K_{1}h^{ab}\right)_{|b}\right]N_{a} +\underbrace{\frac{\sqrt{h}}{16\pi}\left[2\left(K^{ab}-Kh^{ab}\right)N_{a}\right]_{|b}}_{\text{surface term}} -\frac{\sqrt{h}}{16\pi}\left[2\left(\Delta K^{ab}-\Delta Kh^{ab}\right)_{|b}\right]\frac{N_{a}}{N}}\nonumber \\
	& \Scale[0.97]{+\frac{\sqrt{h}}{16\pi}\left[2\left(K_{1}\Delta K^{ab}+\Delta KK_{1}^{ab}\right)\frac{\partial\dot{J}_{ab}}{\partial\dot{h}^{cd}}h^{cd}-\left[2\frac{\partial\dot{J}_{cd}}{\partial\dot{h}^{ab}}+\frac{\partial\dot{J}_{ab}}{\partial\dot{h}^{ef}}\frac{\partial\dot{J}_{cd}}{\partial\dot{h}_{ef}}\right]\left(\Delta K^{cd}K_{1}^{ab}+\Delta K^{ab}K_{1}^{cd}\right)\right]\cdot1}\nonumber \\
	& \Scale[0.97]{+\frac{\sqrt{h}}{16\pi}\left[\left(+h_{ab}\dot{J}^{cd}\frac{\partial\dot{J}_{cd}}{\partial\dot{h}^{ef}}h^{ef}-\dot{J}^{cd}\frac{\partial\dot{J}_{cd}}{\partial\dot{h}^{ab}}-\dot{J}^{cd}\frac{\partial\dot{J}_{ab}}{\partial\dot{h}_{cd}}+\frac{\partial\dot{J}_{ab}}{\partial\dot{h}_{cd}}\dot{h}^{cd}-\dot{J}^{cd}\frac{\partial\dot{J}_{ab}}{\partial\dot{h}_{ef}}\frac{\partial\dot{J}_{cd}}{\partial\dot{h}^{ef}}-2\dot{J}^{ab}\right)K_{1}^{ab}\right]\cdot1}\nonumber\\ 
	&  \Scale[0.97]{-\frac{\sqrt{h}}{16\pi}\left[2K_{1}\Delta K+2\left[K_{1}^{ab}\Delta K_{ab}+5K_{1}\Delta K\right]\right]\cdot1 -\frac{\sqrt{h}}{16\pi}\left[2\Delta K^{2}+\left[\Delta K^{ab}\Delta K_{ab}+5\Delta K^{2}\right]\right]\frac{1}{N}}\nonumber \\
	&  \Scale[0.97]{-\frac{\sqrt{h}}{16\pi}\left[2\Delta K\Delta K^{ab}\frac{\partial\dot{J}_{ab}}{\partial\dot{h}^{cd}}h^{cd}+\left[2\frac{\partial\dot{J}_{cd}}{\partial\dot{h}^{ab}}+\frac{\partial\dot{J}_{ab}}{\partial\dot{h}^{ef}}\frac{\partial\dot{J}_{cd}}{\partial\dot{h}_{ef}}\right]\Delta K^{ab}\Delta K^{cd}\right]\frac{1}{N}}\nonumber \\
	&  \Scale[0.97]{+\frac{\sqrt{h}}{16\pi}\left[2\dot{J}_{ab}\dot{h}^{ab}+2\frac{\partial\dot{J}_{cd}}{\partial\dot{h}_{ab}}\dot{J}^{cd}\dot{h}^{ab}-2\dot{J}^{ab}\dot{J}^{cd}\frac{\partial\dot{J}_{cd}}{\partial\dot{h}^{ab}}-\dot{J}^{cd}\dot{J}^{ij}\frac{\partial\dot{J}_{ij}}{\partial\dot{h}_{ab}}\frac{\partial\dot{J}_{cd}}{\partial\dot{h}^{ab}}\right]\frac{1}{N}}\nonumber \\
	&  \Scale[0.97]{+\frac{\sqrt{h}}{16\pi}\left[\left(h_{ab}\dot{J}^{cd}\frac{\partial\dot{J}_{cd}}{\partial\dot{h}^{ef}}h^{ef}-\dot{J}^{cd}\frac{\partial\dot{J}_{cd}}{\partial\dot{h}^{ab}}-\dot{J}^{cd}\frac{\partial\dot{J}_{ab}}{\partial\dot{h}_{cd}}+\frac{\partial\dot{J}_{ab}}{\partial\dot{h}_{cd}}\dot{h}^{cd}-\dot{J}^{cd}\frac{\partial\dot{J}_{ab}}{\partial\dot{h}_{ef}}\frac{\partial\dot{J}_{cd}}{\partial\dot{h}^{ef}}-2\dot{J}_{ab}\right)\Delta K^{ab}\right]\frac{1}{N}}\label{eq:Mike-vectortensor hamiltonian density}
\end{align}
\end{strip}

From these definitions one can define the extrinsic curvature in term of the new momenta, 
\begin{align*}
	K^{ab} & =\underbrace{\overbrace{\frac{16\pi}{\sqrt{h}}\left[\pi^{ab}-\frac{h^{ab}}{2}\text{Tr}\left\{ \pi^{ab}\right\} \right]}^{\text{Original \ensuremath{K^{ab}}}}+\frac{16\pi}{\sqrt{h}}\frac{\partial\dot{J}_{cd}}{\partial\dot{h}_{ab}}\left[\frac{1}{2}\pi_{J}^{cd}\right]}_{K_{1}^{ab}}\\
	+ &\frac{1}{N}\underbrace{\left[\frac{1}{2}\dot{J}_{ab}-\frac{5}{4}\frac{\partial\dot{J}^{cd}}{\partial\dot{h}_{ab}}\dot{J}_{cd}\right]}_{\Delta K^{ab}} =K_{1}^{ab}+\frac{1}{N}\Delta K^{ab}
\end{align*}

Now $\text{let }\tilde{\pi}=\text{Tr}\left\{ \pi^{ab}\right\} \text{ and }\tilde{\pi}_{J}=h_{ab}\frac{\partial\dot{J}_{cd}}{\partial\dot{h}_{ab}}\pi_{J}^{cd}$.
With this, and a lengthy algebra exercise, the Hamiltonian density can be written as in \eqref{eq:Mike-vectortensor hamiltonian density}. Noticing that 
\begin{equation*}
\left(16\pi/\sqrt{h}\right)^{2}\left[\pi^{ab}\pi_{ab}-\tilde{\pi}^{2}/2\right]-R^{\left(3\right)}
\end{equation*}
contains the original scalar constraint, $-2\left(K_{1}^{ab}-K_{1}h^{ab}\right)_{|b}$ contains the original vector constraint and the notation $A_{|b}$ represents the intrinsic covariant derivative of $A$. We can also note that the surface term now depends on the new degrees of freedom through $K^{ab}$.
\subsection{Further constraints in the deformations}

To guarantee diffeomorphism invariance, variation of the Hamiltonian density with respect to the lapse and shift must vanish, which provide
the constraints. This means that the new vector and scalar constraints are, 
\begin{align*}
	C_{a} & =-\frac{32\pi}{\sqrt{h}}\left(\pi^{ab}+\frac{1}{2}\frac{\partial\dot{J}_{cd}}{\partial\dot{h}_{ab}}\pi_{J}^{cd}-\frac{\tilde{\pi}_{J}}{2}h^{ab}\right)_{|b}
\end{align*}
\begin{align*}
	C_{0} & =\left(\frac{16\pi}{\sqrt{h}}\right)^{2}\left[\pi^{ab}\pi_{ab}-\frac{\tilde{\pi}^{2}}{2}\right]-R^{\left(3\right)}\\
	&+\left(\frac{16\pi}{\sqrt{h}}\right)^{2}\left[\pi_{J}^{ab}\pi_{J,ab}-\frac{\tilde{\pi}_{J}^{2}}{2}\right]-R_{J}^{\left(3\right)}\\
	&+\frac{1}{2}\left(\frac{16\pi}{\sqrt{h}}\right)^{2}\tilde{\pi}\tilde{\pi}_{J}-V_{J}\left(J,h\right)\\
	& +2K_{1}K_{1}^{ab}\frac{\partial\dot{J}_{ab}}{\partial\dot{h}^{cd}}h^{cd}-K_{1}^{ab}K_{1ab}-5K_{1}^{2}\\
	&-\left[2\frac{\partial\dot{J}_{cd}}{\partial\dot{h}^{ab}}+\frac{\partial\dot{J}_{ab}}{\partial\dot{h}^{ef}}\frac{\partial\dot{J}_{cd}}{\partial\dot{h}_{ef}}\right]K_{1}^{cd}K_{1}^{ab}
\end{align*}

Furthermore, the new Hamiltonian density \eqref{eq:Mike-vectortensor hamiltonian density} contains terms proportional to $\frac{1}{N}$,$\frac{N_{1}}{N}$ and $1$. In order for Hamilton's equations to not depend on a gauge choice, these terms must individually vanish. The terms proportional to $\frac{1}{N}$,$\frac{N_{a}}{N}$, which will be called $C_{-1}$ and $C_{-1}^{a}$ respectively, must
vanish separately so as to not make the vector and scalar constraints depend on the gauge, a crucial feature for any gauge theory. The term proportional to unity, let us call it $C_{1}$, represents a type of \textit{bare} Hamiltonian that must also vanish independently. If this term were to persist, then within Hamilton's equations there would exist a gauge choice of $\left(N,N^{a}\right)$ such that there is no time
evolution, which would allow there to be a universal reference frame. In both cases, if these terms do not vanish, diffeomorphism invariance
will be broken and hence\footnote{See \url{https://github.com/josegalvez/HD-GR} for a generalized step-by-step explanation of all the calculations shown this subsection.}, 
\begin{align*}
	C_{1} & =2\left(K_{1}\Delta K^{ab}+\Delta KK_{1}^{ab}\right)\frac{\partial\dot{J}_{ab}}{\partial\dot{h}^{cd}}h^{cd}-\\&\left[2\frac{\partial\dot{J}_{cd}}{\partial\dot{h}^{ab}}+\frac{\partial\dot{J}_{ab}}{\partial\dot{h}^{ef}}\frac{\partial\dot{J}_{cd}}{\partial\dot{h}_{ef}}\right]\left(\Delta K^{cd}K_{1}^{ab}+\Delta K^{ab}K_{1}^{cd}\right)\\
	& + \bigg(h_{ab}\dot{J}^{cd}\frac{\partial\dot{J}_{cd}}{\partial\dot{h}^{ef}}h^{ef}-\dot{J}^{cd}\frac{\partial\dot{J}_{cd}}{\partial\dot{h}^{ab}}-\dot{J}^{cd}\frac{\partial\dot{J}_{ab}}{\partial\dot{h}_{cd}}\\
	& +\frac{\partial\dot{J}_{ab}}{\partial\dot{h}_{cd}}\dot{h}^{cd}- \dot{J}^{cd}\frac{\partial\dot{J}_{ab}}{\partial\dot{h}_{ef}}\frac{\partial\dot{J}_{cd}}{\partial\dot{h}^{ef}}-2\dot{J}^{ab}\bigg)K_{1}^{ab}\\
	& -\left(2K_{1}\Delta K+2\left[K_{1}^{ab}\Delta K_{ab}+5K_{1}\Delta K\right]\right)
\end{align*}

\[
C_{-1}^{a}=2\left(\Delta K^{ab}-\Delta Kh^{ab}\right)_{|b}
\]
\begin{align*}
	C_{-1} & =-\bigg(2\Delta K\Delta K^{ab}\frac{\partial\dot{J}_{ab}}{\partial\dot{h}^{cd}}h^{cd}+\\
	&\left[2\frac{\partial\dot{J}_{cd}}{\partial\dot{h}^{ab}}+\frac{\partial\dot{J}_{ab}}{\partial\dot{h}^{ef}}\frac{\partial\dot{J}_{cd}}{\partial\dot{h}_{ef}}\right]\Delta K^{ab}\Delta K^{cd}\bigg)\\
	& -\left(2\Delta K^{2}+\left[\Delta K^{ab}\Delta K_{ab}+5\Delta K^{2}\right]\right)\\
	& +\bigg(2\dot{J}_{ab}\dot{h}^{ab}+2\frac{\partial\dot{J}_{cd}}{\partial\dot{h}_{ab}}\dot{J}^{cd}\dot{h}^{ab}-2\dot{J}^{ab}\dot{J}^{cd}\frac{\partial\dot{J}_{cd}}{\partial\dot{h}^{ab}}\\
	&-\dot{J}^{cd}\dot{J}^{ij}\frac{\partial\dot{J}_{ij}}{\partial\dot{h}_{ab}}\frac{\partial\dot{J}_{cd}}{\partial\dot{h}^{ab}}\bigg)\\
	& +\bigg(h_{ab}\dot{J}^{cd}\frac{\partial\dot{J}_{cd}}{\partial\dot{h}^{ef}}h^{ef}-\dot{J}^{cd}\frac{\partial\dot{J}_{cd}}{\partial\dot{h}^{ab}}-\dot{J}^{cd}\frac{\partial\dot{J}_{ab}}{\partial\dot{h}_{cd}}\\
	&+\frac{\partial\dot{J}_{ab}}{\partial\dot{h}_{cd}}\dot{h}^{cd}-\dot{J}^{cd}\frac{\partial\dot{J}_{ab}}{\partial\dot{h}_{ef}}\frac{\partial\dot{J}_{cd}}{\partial\dot{h}^{ef}}-2\dot{J}_{ab}\bigg)\Delta K^{ab}
\end{align*}

\section{Deforming general relativity in tetrad theory and Ashtekar variables}\label{sec:Ashtekar}
There exist other representations of general relativity aside from that presented above. Here we will briefly review tetrad formalisms. In covariant form arguably the most popular version of a tetrad action is the tetrad-Palatini action \cite{ref:tetpalat} which may be written, without the Holst term for simplicity, as:
\begin{align}
S&=\int \dform{e}\wedge \dform{e} \wedge \starf(\dform{\omega}) \nonumber \\
& =\frac{1}{2}\int d^{4}x\, \epsilon_{IJKL} \epsilon^{\mu\nu\rho\sigma} e^{I}_{\;\mu} e^{J}_{\;\nu} F^{KL}_{\;\;\rho\sigma}\,, \label{eq:tpact}
\end{align}
with $e$ the tetrad and $\omega$ a Lorentz (for our purposes) connection whose dual (on the capital indices, coupling to the $so(3,1)$ algebra) field-strength is denoted by $\starf$. $e$ and $\omega$ constitute the independent fields and the variation with respect to the connection yields the torsionless condition for the connection whereas variation with respect to the tetrad yields the Einstein equations. The configuration space is spanned by the pair of components $(e^{I}_{\;\mu},\; \omega^{JK}_{\;\;\;\;\nu})$.

Let us now consider deformations of $e$ and $\omega$ in the action (\ref{eq:tpact}) while noting that the tetrad $e$ should remain one of the variables in the new theory (analogous to keeping the metric variable in the deformed metric theory). Using slightly modified notation we write
\begin{equation}
 \dform{e} \rightarrow \mathcal{E}(\te,\tw)\,,\; \tw \rightarrow \mathcal{W}(\te,\tw)\,. \label{eq:tpdeform}
\end{equation}
Note that this is equivalent to the deformation
\begin{equation}
 \dform{e} \rightarrow \te + \Delta \te\,,\; \dform{\omega} \rightarrow \tw + \Delta\tw \label{eq:tpdeltadeform}
\end{equation}
with
\begin{equation}
 \Delta \te := \mathcal{E}(\te,\tw) -\te \,,\; \Delta \tw := \mathcal{W} -\tw\,.
\end{equation}
As a clarification, what we mean by this is that the components of $\dform{e}$ and $\dform{\omega}$, that is $e^{I}_{\;\mu}$ and $\tilde{\omega}^{IJ}_{\;\;\;\mu}$, are the quantities deformed. Of course the group generator and differential form structure remain unaltered.

Using (\ref{eq:tpdeltadeform}) in (\ref{eq:tpact}), and noting that $\tilde{\dform{F}}=\mathrm{d}\tw+\tw\wedge\tw$, we define the quantity
\begin{equation}
\Scale[0.94]{ \Delta \starftilde:={^{*}\left[\mathrm{d}(\Delta \tw) + \tw\wedge \Delta \tw + \Delta\tw\wedge \tw +\Delta\tw\wedge \Delta \tw\right]}\,.} \label{eq:deltaF}
\end{equation}
Noting that the action implicitly involves a trace, the resulting action may be written as:
\begin{align}
S^{\prime}=S+\int& \Big\{2\left[\te\wedge \Delta \te \wedge \starf + \te\wedge \Delta \te \wedge \Delta\starf\right]  \nonumber \\
&+ \te\wedge \te \wedge \Delta \starf + \Delta \te\wedge \Delta \te \wedge \starf  \nonumber\\
&+  \Delta \te \wedge \Delta \te \wedge \Delta\starf \Big\}\,. \label{eq:tetsprime}
\end{align}
Deformations of the tetrad-Palatini action (\ref{eq:tpact}) are particularly straight-forward to reproject via (\ref{decompose}). In four spacetime dimensions the manifold of 4-forms is one-dimensional, and hence we may write \footnote{The numerical structure of the three-index permutation symbols   $\epsilon^{\nu\rho\sigma}$ is implied by equation (\ref{eq:palatdecompose}). It is a different quantity for each value of the index $\mu$ in (\ref{eq:palatdecompose}). The antisymmetrization is to preserve the algebraic structure of the four index Levi-Civita when dealing with components, as we are here.}
\begin{equation}
\epsilon^{\mu\nu\rho\sigma} \propto N^{[\mu}\epsilon^{\nu\rho\sigma]}\,. \label{eq:palatdecompose}
\end{equation}
Therefore the modified action can be cast as the original action plus a term of the form
\begin{equation}
\Delta S \propto \int d^{4}x \, \epsilon^{\nu\rho\sigma}N^{\mu} \Delta \mathcal{L}^{\prime}_{\mu\nu\rho\sigma}
\end{equation}
where $\Delta\mathcal{L}^{\prime}_{\mu\nu\rho\sigma}$ is the deformation of the
\begin{equation}
\epsilon_{IJKL} e^{I}_{\;\mu} e^{J}_{\;\nu} F^{KL}_{\;\;\rho\sigma}\, \nonumber
\end{equation}
part in the Lagrangian. The antisymmetric structure of the Levi-Civita (written as in (\ref{eq:palatdecompose})) along with its contraction with the tetrads in the undeformed action (\ref{eq:tpact}) is actually sufficient to filter out terms linear proportional to $N$ and $N^{a}$. However, for a generic deformation, the new action may need to be reprojected in the manner illustrated in the previous sections. Simple deformations (for example those that do not alter the linear dependence on $N^{\mu}$ in the new variables) will not suffer from this.

Before proceeding we should note that $\Delta \tw$ may depend on the tetrad and, as mentioned, we wish to retain the tetrad as a fundamental degree of freedom in the new theory. Therefore we can state the following points about variable deformations in the Palatini Lagrangian:
\begin{itemize}
 \item Since $\mathrm{d}(\Delta \tw)=\mathrm{d}\mathcal{W}(\te,\tw)-\mathrm{d}\tw$, in general we pick up a generalized ``velocity'' conjugate to the tetrad as now derivatives of the tetrad are explicitly present in the action. This is an example of acquiring new degrees of freedom via the deformation.
 \item One needs to identify the corresponding connection variable(s). These other variable(s) may or may not comprise $\tilde{\omega}$. The situation regarding the new variables tends to be clearer in the Hamiltonian picture, which is the main focus of this paper, and will be discussed below when we go to the Hamiltonian formalism.
 \item The new connection may no longer be torsion free.
 \item Related to the previous point, the tetrad's covariant derivative may no longer be annihilated. This is not surprising since we now have a new connection, in a new theory whose symmetry may differ from the original symmetry. The new constraints will serve to enforce this new symmetry.  
\end{itemize}

Next we wish to address the main point, which is the issue of Hamiltonian consistency of a deformed theory. As in the introduction, we shall first consider the undeformed action in its canonical guise, then identify the configuration-momentum variables, and then perform the variable deformation. Since we will be deforming the previous action, but in a different form and also a different set of variables, the deformed theory will not necessarily be the same as if one deforms the Palatini action directly as above. (This is also true due to discarding surface terms when transforming one undeformed action into another form of the same action.) The tetrad-Palatini action (with the Holst term) in canonical form leads almost directly to the Ashtekar-Barbero action. The transformations required from the tetrad-Palatini action to the Ashtekar action may be found in \cite{ref:AEdotact}. The Ashtekar variables traditionally comprise an $su(2)$ valued densitized triad, $E_{i}^{\;a}=\mbox{det}|e|e_{i}^{\;a}$, and a connection, $A^{i}_{\;a}$, where the indices $i,\,j,$ etc. couple to the $SU(2)$ generators (to use notation most often seen in loop quantum gravity).  The relationships between these new variables and the ADM variables are given by
\begin{subequations}
\romansubs
{\allowdisplaybreaks\begin{align}
  h h^{ab}=&E_{i}^{\;a}E_{j}^{\;b} \delta^{ij}\,, \label{eq:Ehreln} \\
  A^{i}_{\;a}=&\Gamma^{i}_{\;a}+\gamma K^{i}_{\;a}\,. \label{eq:AKreln}
 \end{align}}
\end{subequations}
\noindent Here $\Gamma^{i}_{\;a}$ is the spin connection, annihilating an orthonormal triad via covariant differentiation, $K^{i}_{\;a}$ is the densitized extrinsic curvature
\begin{equation}
 K^{i}_{\;a}:=\frac{1}{\sqrt{E}}K_{ab}E_{j}^{\;b}\delta^{ij}, \label{eq:densitizedext}
\end{equation}
and $h$ is $\det{h_{ab}}$. The quantity $\gamma$ is known as the Barbero-Immirzi parameter, which from the point of view of the bulk classical equations of motion is arbitrary, but its exact value is of importance in the quantum theory. It has been shown that the variables $E_{i}^{\;a}$, $A^{i}_{\;a}$ are related to the ADM variables in the previous section via a canonical transformation \cite{ref:admtoash1, ref:admtoash2}. 

The Poisson algebra in these variables is given by the brackets
\begin{subequations}
\romansubs
{\allowdisplaybreaks\begin{align}
  \left\{E_{i}^{\;a}(x),\,A^{j}_{\;b}(y)\right\}=& \kappa \gamma\delta(x,y)\delta_{i}^{\;j} \delta^{a}_{b},\, \label{eq:ashpois1} \\
  \left\{E_{i}^{\;a}(x),\,E_{j}^{\;b}(y)\right\}=& \left\{A^{i}_{\;a}(x),\,A^{j}_{\;b}(y)\right\}=0\,. \label{eq:ashpois2}
\end{align}}
\end{subequations}
In these variables the action may be written as 
\begin{equation}
\Scale[0.84]{ S=\frac{1}{\kappa}\bigintss dt \bigintss d^{3}x\left[E_{i}^{\;a} \dot{A}^{i}_{\;a} - \mathcal{H}(N) - \mathcal{H}(N^{a}) - \mathcal{G}(\lambda^{i})\right]\,} \label{eq:ashact}
\end{equation}
plus corresponding possible matter terms. In this case the scalar and vector constraints are given by
\begin{subequations}
\romansubs
{\allowdisplaybreaks\begin{align}
\mathcal{H}(N)=& \Scale[0.95]{N \frac{E_{i}^{\;a}E_{j}^{\;b}}{\sqrt{E}} \left(F^{k}_{\;ab} \epsilon^{ij}_{\;\;\;k}-2(\gamma^{2}-1)K^{i}_{\;[a}K^{j}_{\;b]}\right)\,,} \label{eq:ashscal}\\                   
\mathcal{H}(N^{b})=& N^{b}\left[E_{i}^{\;a}F^{i}_{\;ab}-(\gamma^{2}+1)K^{i}_{\;b}G_{i}\right] \,. \label{eq:ashvec}
\end{align}}
\end{subequations}
$F^{i}_{\;ab}$ is the field-strength tensor of the connection, $F^{i}_{\;ab}:=\partial_{a}A^{i}_{\;b}- \partial_{b}A^{i}_{\;a}+\epsilon^{i}_{\;jk}A^{j}_{\;a}A^{k}_{\;b}$. Note that since we are dealing with a (densitized) triad variable in lieu of a metric variable, a new constraint is introduced which fixes the internal $SU(2)$ rotation of the triad. This is the so-called Gauss constraint,
\begin{equation}
G_{i}\lambda^{i}=:\mathcal{G}(\lambda^{i}):=\left(\partial_{a} E_{i}^{\;a} + \epsilon_{ij}^{\;\;k} A^{j}_{\;a} E_{k}^{\;a}\right) \lambda^{i}\,, \label{eq:gaussconst}
\end{equation}
with its own Lagrange multiplier, $\lambda^{i}$, which fixes the metricity condition on the densitized triad. 

Since the internal spatial geometry is encoded in the densitized triad we will, in analogy to keeping the 3-metric as the configuration variable in the ADM variables, keep the densitized triad as the canonical gravitational momentum variable after deformation. In brief, as in the previous section the deformation may be written schematically as
\begin{subequations}
\romansubs
{\allowdisplaybreaks\begin{align}
E_{i}^{\;a} \rightarrow \tilde{E}_{i}^{\;a}+\Delta\tilde{E}_{i}^{\;a}\,, \label{eq:Edeform}\\
A^{i}_{\;a} \rightarrow \tilde{A}^{i}_{\;a}+\Delta\tilde{A}^{i}_{\;a}\,. \label{eq:Adeform}
\end{align}}
\end{subequations}

Before continuing, it should be noted that in principle the quantities in the Gauss constraint are to be deformed so that the Gauss constraint becomes some complicated function of both the densitized triad and the connection. However, caution should be applied in this case since the Gauss constraint, in its original form, enforces the specific condition of metricity fixing. It may be desirable, regardless of the specific deformation, to enforce ``by hand'' that the Gauss constraint transform as $\mathcal{G}(\lambda_{i}, E_{j}^{\;a}) \rightarrow \mathcal{G}(\lambda_{i}, \tilde{E}_{j}^{\;a})$ so that after deformation triad metricity fixing is still enforced by this constraint. A comment on this follows the example below.

The general deformation of the variables (\ref{eq:Edeform}) and (\ref{eq:Adeform}) be applied to (\ref{eq:ashact}) and the resulting action re-written in Hamiltonian canonical form by identifying the new configuration-momentum pairs. In principle this may be done, but the resulting action will generally be very complicated and not very perspicuous, not least because of the fact that $K^{i}_{\;a}=\gamma^{-1}[A^{i}_{\;a}-\Gamma^{i}_{\;a}]$ and the spin connection is given by the complicated expression
\begin{align}
 \Gamma^{i}_{\;a}=&\frac{1}{2}\epsilon^{i\;\;k}_{\;j} E_{k}^{\;b}\left[-2\partial_{[a}E^{j}_{\;b]} + E^{j}_{\;c} E^{\ell}_{\;a} \partial_{b}E_{\ell}^{\; c}\right]\nonumber \\
 & +\frac{1}{4}\epsilon^{i\;\;k}_{\;j} E_{k}^{\;b}\left[ 2 E^{j}_{\;a} \partial_{b} \ln(E) - E^{j}_{\;b} \partial_{a} \ln(E)\right]\,,
\end{align}
as well as the contracted triad $E_{j}^{\;a}=E_{k}^{\;a}E_{\ell}^{\;b} \delta^{kl} E^{i}_{\;b} \delta_{ij}$. Therefore we illustrate the scheme on an example relevant to some studies of loop quantum gravity. A specific example of such deformations in these variables is provided by the often used holonomy correction inspired by loop quantum gravity. In such scenarios the connection variable is deformed in order to represent loop quantum corrections of the connection as a holonomy; namely, in the language of (\ref{ct1}) and (\ref{ct2}):
\begin{subequations}
\romansubs
{\allowdisplaybreaks\begin{align}
 A^{i}_{\;a}\rightarrow& H^{i}_{\;a}\left(\tilde{A}^{i}_{\;a}, \tilde{E}_{j}^{\;b}\right)= 
 \frac{\sin\left[\tilde{A}^{i}_{\;a}\,\delta(\tilde{E}_{j}^{\;b})\right]}{\delta(\tilde{E}_{j}^{\;b})}\,,\label{eq:hocor1}\\
E_{i}^{\;a} \rightarrow& P_{i}^{\;a}\left(\tilde{A}^{i}_{\;a}, \tilde{E}_{j}^{\;b}\right) = \tilde{E}_{i}^{\;a}\,, \label{eq:holcor2} \\
\Delta \tilde{A}^{i}_{\;a}=&\frac{\sin\left[\tilde{A}^{i}_{\;a}\,\delta(\tilde{E}_{j}^{\;b})\right]}{\delta(\tilde{E}_{j}^{\;b})}-\tilde{A}^{i}_{\;a}\,,\;\Delta \tilde{E}_{i}^{\;a} =0\,. \label{eq:holcor3}
\end{align}}
\end{subequations}
The quantity $\delta(\tilde{E}_{j}^{\;b})$ is related to the proper-length along the path which the holonomy is taken, and hence depends on $\tilde{E}_{j}^{\;b}$ but not the connection \cite{ref:ashsingh} (although sometimes $\delta$ is taken to be constant for simplicity). Note that here the deformation is performed at the level of the Hamiltonian variables, compatible with the analysis of this manuscript. This is due to the fact that the holonomy correction is inspired by the operator representation of the algebra of the commutator of loop quantum gravity, and therefore these modifications are often applied via direct substitution of (\ref{eq:hocor1}) and (\ref{eq:holcor2}) into the action (\ref{eq:ashact}). The issue now remains as to what to do with the canonical term in (\ref{eq:ashact}), $E_{i}^{\;a} \dot{A}^{i}_{\;a}$. One method is to simply insert the time derivative of (\ref{eq:Adeform}) in lieu of $\dot{A}_{i}^{\;a}$\,.
{\emph{If}} one proceeds in this manner the resulting action is
\begin{align}
\displaystyle
S^{\prime}=&\Scale[0.90]{\frac{1}{\kappa} \bigintss dt \bigintss d^{3}x \left[P_{i}^{\;a} \dot{H}^{i}_{\;a} - \tilde{\mathcal{H}}(N) -\tilde{\mathcal{H}}(N^{b}) - \tilde{\mathcal{G}}(\lambda^{i})\right]}\nonumber \\
=&\frac{1}{\kappa} \int dt \int d^{3}x\, \bigg\{\tilde{E}_{i}^{\;a} \bigg[\cos(\tilde{A}^{i}_{\;a} \delta) \dot{\tilde{A}}^{i}_{\;a} \nonumber \\
+ & \left(\cos(\tilde{A}^{i}_{\;a} \delta) \frac{\partial \delta}{\partial \tilde{E}_{j}^{\;b}} \delta^{-1} - \sin(\tilde{A}^{i}_{\;a} \delta)  \frac{\partial \delta}{\partial \tilde{E}_{j}^{\;b}} \delta^{-2}\right) \dot{\tilde{E}}_{j}^{\;b}\bigg] \nonumber \\
- &  \hat{\mathcal{H}}(N) -\hat{H}(N^{b}) - \hat{G}(\lambda^{i})\bigg\}\,. \label{eq:holdefact}
\end{align}
(Due to the nonlinear nature of the deformation, the index structure is awkward in (\ref{eq:holdefact}). In each term repeated indices are summed, even if they appear more than twice.)
The natural interpretation of the above is that the resulting theory has more degrees of freedom than general relativity; the Hamiltonian degrees of freedom being:
\begin{subequations}
\romansubs
{\allowdisplaybreaks
 \begin{align}
  \tilde{A}^{i}_{\;a},\;& \Scale[0.95]{\mbox{\tiny{$_{A}$}}\tilde{\pi}_{i}^{\;a}:=\tilde{E}_{i}^{\;a}\cos(\tilde{A}^{i}_{\;a} \delta)}\;\; \mbox{(no sum)} \label{eq:defashvars1}, \\[0.1cm]
  \mbox{and}& \nonumber \\
  \tilde{E}_{i}^{\;a},\;& \Scale[0.90]{\mbox{\tiny{$_{E}$}}\tilde{\pi}^{j}_{\;b}:=\tilde{E}_{i}^{\;a}\left(\cos(\tilde{A}^{i}_{\;a} \delta) \frac{\partial \delta}{\partial \tilde{E}_{j}^{\;b}} \delta^{-1} - \sin(\tilde{A}^{i}_{\;a} \delta)  \frac{\partial \delta}{\partial \tilde{E}_{j}^{\;b}} \delta^{-2}\right)}. \label{eq:defashvars2}
 \end{align}}
\end{subequations}
It is noted that in this particular scenario the system retains second-order equations of motion, and therefore the Ostrogradskii stability issues presented in the previous section are avoided. However, if one a priori assumes some relationship between $\tilde{A}^{i}_{\;a}$ and $\tilde{E}_{i}^{\;a}$ (such as in, for example, electromagnetism where the electric field is related to the time derivative of the potential), then one could induce a higher-order theory via the $\dot{\tilde{E}}_{j}^{\;b}$ term in (\ref{eq:holdefact}). We do not assume this since the canonical degrees of freedom are independent here.

The Poisson brackets are now defined with respect to the new degree of freedom variables. That is
\begin{align}
 \left\{X,\,Y\right\} = & \frac{\partial X}{\partial \tilde{A}^{i}_{\;a}} \frac{\partial Y}{\partial (\mbox{\tiny{$_{A}$}}\tilde{\pi}_{i}^{\;a})} -  \frac{\partial X}{\partial(\mbox{\tiny{$_{A}$}}\tilde{\pi}_{i}^{\;a})} \frac{\partial Y}{\partial \tilde{A}^{i}_{\;a} } \nonumber \\
 & +\frac{\partial X}{\partial \tilde{E}_{i}^{\;a}} \frac{\partial Y}{\partial (\mbox{\tiny{$_{E}$}}\tilde{\pi}^{i}_{\;a})} -  \frac{\partial X}{\partial(\mbox{\tiny{$_{E}$}}\tilde{\pi}^{i}_{\;a})} \frac{\partial Y}{\partial \tilde{E}_{i}^{\;a} }\,. \label{eq:holonomypoisson}
\end{align}

The decomposition (\ref{decompose}) is straight-forward since the deformation of $\tilde{\mathcal{H}}(N)$ is still only proportional to $N$ and the deformation of $\tilde{\mathcal{H}}(N^{a})$ is still only proportional to $N^{a}$. This results in the following constraints in the new theory:
\begin{subequations}
\romansubs
{\allowdisplaybreaks\begin{align}
\Scale[0.93]{\hat{\mathcal{H}}(N)=}&\Scale[0.93]{N\frac{\tilde{E}_{i}^{\;a} \tilde{E}_{j}^{\;b}}{\sqrt{\tilde{E}}}\Big\{ \left(\underset{o}{\tilde{F}}^{k}_{\;ab} + \Delta \tilde{F}^{k}_{\;ab}\right)\epsilon^{ij}_{\;\;\;k} } \label{eq:newashscal}\\
&\Scale[0.95]{-2(\gamma^{2}-1) \left(\underset{o}{\tilde{K}}+\Delta \tilde{K}\right)^{i}_{\;[a} \left(\underset{o}{\tilde{K}}+\Delta \tilde{K}\right)^{j}_{\;b]} \Big\} \,,} \nonumber \\[0.2cm]
\Scale[0.90]{\hat{\mathcal{H}}(N^{b})=}&\Scale[0.90]{N^{b}\Big\{\tilde{E}_{i}^{\;a} \left[\underset{o}{\tilde{F}}^{i}_{\;ab} +\Delta \tilde{F}^{i}_{\;ab} \right]  -(\gamma^{2}+1) \left[ \underset{o}{\tilde{K}}^{i}_{\;b} \tilde{G}_{i} \right.} \nonumber \\
&\Scale[0.90]{ + \left. \underset{o}{\tilde{K}}^{i}_{\;b} \Delta \tilde{G}_{i} + \Delta \tilde{K}^{i}_{\;b} \underset{o}{\tilde{G}}_{i} + \Delta \tilde{K}^{i}_{\;b}\Delta \tilde{G}_{i}\right]\Big\}\,,} \label{eq:newashdif} \\[0.2cm]
\Scale[0.93]{\hat{\mathcal{G}}(\lambda^{i}) = }&\Scale[0.93]{\lambda^{i}\left[\underset{o}{\tilde{G}}_{i} + \Delta \tilde{G}_{i}\right]\,.} \label{eq:newashgauss}
\end{align}}
\end{subequations}
Here, the $o$ subscript refers to the original (pre\ -\ deformation) quantities and the remaining quantities are as follows:
\begin{subequations}
\romansubs
{\allowdisplaybreaks\begin{align}
 \Scale[0.94]{\Delta \tilde{F}^{i}_{\;ab}:=}&\Scale[0.85]{2\partial_{[a}(\Delta \tilde{A}^{i}_{\;b]}) +\epsilon^{i}_{\;jk}\left[\tilde{A}^{j}_{\;a} \Delta \tilde{A}^{k}_{\;b} + \Delta \tilde{A}^{j}_{\;a} \tilde{A}^{k}_{\;b} +\Delta \tilde{A}^{j}_{\;a} \Delta \tilde{A}^{k}_{\;b} \right] \, \label{eq:ashdeltaF}} \\
 {\Delta \tilde{K}^{i}_{\;a}:=}& \frac{1}{\gamma} \Delta \tilde{A}^{i}_{\;a} \, \label{eq:ashdeltaK} \\
 {\Delta \tilde{G}_{i}:=}& \epsilon_{ij}^{\;\;\;k} \tilde{E}^{a}_{\;k} \Delta \tilde{A}^{j}_{\;a}\,. \label{eq:ashdeltaG}
\end{align}}
\end{subequations}
Note specifically that the deformed Gauss constraint, $\hat{\mathcal{G}}(\lambda^{i})$, can be satisfied either by making terms in $\Delta G_{i}$ cancel terms in $\underset{o}G_{i}$, or else by enforcing $\underset{o}G_{i}$ and $\Delta G_{i}$ equal to zero separately. In the first case the Gauss constraint no longer enforces metricity fixing of the densitized tetrad whereas in the second case it does (since the Gauss constraint in its original form vanishes) and hence is generally preferable. In the second case there will be the enforcement of the metricity fixing condition along with another constraint demanded by the vanishing of (\ref{eq:ashdeltaG}).

The action (\ref{eq:holdefact}) along with the set of constraints (\ref{eq:newashscal}), (\ref{eq:newashdif}) (supplemented with (\ref{eq:newashgauss})) yields the deformed theory whose constraints will obey the constraint algebra (\ref{constraintalg1}-{}-\ref{constraintalg3}). However, although this scenario was reasonably straight-forward to implement, it should be noted that the variables which undergo Hamiltonian evolution in this example are not $A^{i}_{\;b}$ and $E_{j}^{\;b}$ but the configuration and momentum variables in (\ref{eq:defashvars1}) and (\ref{eq:defashvars2}). The deformed theory must be written in terms of these variables, which depends on the invertibility of the transformation equations (\ref{eq:defashvars1}) and (\ref{eq:defashvars2}). One must therefore know the explicit form of $\delta(E_{j}^{\;b})$ in order to do this. In other words, the equation (\ref{eq:variation2}), utilized in order to derive the algebra among the constraints, picks up the extra degrees of freedom:
\begin{align}
\delta_\zeta q\frac{\delta I}{\delta q}= & \delta_\zeta \tilde{A}^{i}_{\;a}\frac{\delta I}{\delta \tilde{A}^{i}_{\;a}}+\delta_\zeta (\piA{i}{a})\frac{\delta I}{\delta (\piA{i}{a})} \nonumber \\
&+\delta_\zeta \tilde{E}_{i}^{\;a}\frac{\delta I}{\delta \tilde{E}_{i}^{\;a}}+\delta_\zeta (\piE{i}{a})\frac{\delta I}{\delta (\piE{i}{a})}, \label{eq:ashvariation}
\end{align}
where here we have only considered the gravitational sector. It is important to note at this stage that it is generally a non-trivial matter to rewrite the action, and resulting equations of motion, solely in terms of the new variables. It is also possible, depending on the particular transformation, that there is no unique way of writing the deformed action in terms of the new variables. In the case of the holonomy corrections presented above, \emph{one} way to proceed is by simply leaving the action in the form (\ref{eq:holdefact}), since $A^{i}_{\;a}$ and $E_{i}^{\;a}$ are legitimate degrees of freedom even in the deformed theory as can be seen from (\ref{eq:defashvars1}-ii). That would however lead to a theory which would only contain stationary gravity ($\dot{\tilde{E}}_{i}^{\;a}=\partial \mathcal{H}/\partial \mbox{\tiny{$_{E}$}}\tilde{\pi}^{i}_{\;a}=0$,  $\dot{\tilde{A}}^{i}_{\;a}=\partial \mathcal{H} / \partial{\mbox{\tiny{$_{A}$}}\tilde{\pi}_{i}^{\;a}}$=0) as there is no dependence on either of the canonical momenta. In fact, all Poisson brackets between the constraints would also trivially vanish. (Convexity would also be lost therefore introducing ambiguity in transforming to the Lagrangian formulation.)

There is also another form of the canonical form of the action, which essentially reverses the roles of $A$ and $E$ as configuration and momentum \cite{ref:AEdotact}. In this guise the gravitational action has the canonical form
\begin{equation}
\Scale[0.84]{ S=\frac{1}{\kappa}\bigintss dt \bigintss d^{3}x\left[{A}^{i}_{\;a}\dot{E}_{i}^{\;a} + \mathcal{H}(N) + \mathcal{H}(N^{a}) + \mathcal{G}(\lambda^{i})\right]\,.} \label{eq:aedotact}
\end{equation}
In order to employ a straight-forward holonomy correction to this form of the action one must utilize the momentum representation of the quantum operators \cite{ref:momrep}. Here the holonomy correction for the momentum representation takes the form
\begin{subequations}
\romansubs
{\allowdisplaybreaks\begin{align}
 A^{i}_{\;a}\rightarrow& P^{i}_{\;a}\left(\tilde{A}^{i}_{\;a}, \tilde{E}_{j}^{\;b}\right)= 
 -\frac{\sin\left[\tilde{A}^{i}_{\;a}\,\delta(\tilde{E}_{j}^{\;b})\right]}{\delta(\tilde{E}_{j}^{\;b})}\,,\label{eq:momholcor1}\\
E_{i}^{\;a} \rightarrow& H_{i}^{\;a}\left(\tilde{A}^{i}_{\;a}, \tilde{E}_{j}^{\;b}\right) = \tilde{E}_{i}^{\;a}\,, \label{eq:momholcor2}
\end{align}}
\end{subequations}
which result in the corresponding modified action
\begin{align}
 S^{\prime}=&\frac{1}{\kappa}\int dt \int d^{3}x\left[-\frac{\sin\left(\tilde{A}^{i}_{\;a}\delta(\tilde{E}_{j}^{\;b})\right)}{\delta(\tilde{E}_{j}^{\;b})}\dot{\tilde{E}}_{i}^{\;a}\right. \nonumber \\
&\left.  + \hat{\mathcal{H}}(N) + \hat{\mathcal{H}}(N^{a}) + \hat{\mathcal{G}}(\lambda^{i})\right]\,. \label{eq:aedotactprime}
\end{align}
Note that here, on the other hand, the corrected theory retains the same number of degrees of freedom as the original, though they are not the same ones as in the original theory:
\begin{equation}
\tilde{E}_{i}^{\;a},\;\;\; \mbox{\tiny{$_{E}$}}\tilde{\pi}^{i}_{\;a}:=-\frac{\sin\left(\tilde{A}_{i}^{\;a}\delta(\tilde{E}_{j}^{\;b})\right)}{\delta(\tilde{E}_{j}^{\;b})}\,.
\end{equation}
In this case the resulting action can certainly be written in terms of the new canonical variables only. We will not explicitly calculate the constraints in this case as they are more straight-forward here than in the previous scenario. It is again important to note that the same base theory, written in different guises, gives rise to completely different theories under variable deformations. It might seem that in this second case one has not gained a new theory, since the new phase-space variables, $\tilde{E}_{i}^{\;a}$ and $\mbox{\tiny{$_{E}$}}\tilde{\pi}^{i}_{\;a}$, evolve in exactly the same way as the old ones, ${E}_{i}^{\;a}$ and ${A}^{i}_{\;a}$, under Hamiltonian evolution. However, information about the geometry is encoded in the metric (or ${E}_{i}^{\;a}$) and the extrinsic curvature (related to ${A}^{i}_{\;a}$ via (\ref{eq:AKreln}) ). Therefore, even though the new variables evolve in the same way as the old, the \emph{geometry} will not evolve in the same way as in the undeformed theory.

We end this example with a comment that the above deformation method is not the usual way that holonomy corrections are implemented. This is because in the usual scheme one wishes to demand that the connection and densitized tetrad remain canonically conjugate variables of the theory as in (\ref{eq:ashpois1},ii). However, it has been shown that in certain cases that may lead to inconsistencies in the Poisson algebra of the constraints when matter is present \cite{Bojowald:2015zha}. The scheme presented in this manuscript is specifically designed to ensure consistency of the algebra even in the presence of matter.

\section{Concluding remarks}
In this paper, we proposed a systematic treatment of theories generated by the modification of the canonical variables of general relativity both in metric variables and tetrad variables. The objective of this approach is to explore under which conditions the constraint algebra retains its first-class structure (which supports stable gauge fixing) and its diffeomorphic symmetries. The deformed fields and momenta were re-introduced in the original theory in order to observe the effects of the transformed variables. We covered two possible cases (I) deformations that do not introduce new higher-derivatives terms in the action and (II) non-unitary modifications that will create new degrees of freedom. Furthermore, we evaluated the transformed gravitational variables in two well-known incarnations of the gravitational action: the Einstein-Hilbert action and in the tetrad formalism of the Palatini action and in the Ashtekar variables. We must remark, however, that it is viable to deform the canonical variables of other actions following the same procedure we described in this paper. 

In the case of the variable deformations in the Einstein-Hilbert action, the preservation of the original number of degrees of freedom in case (I) unavoidably leads us into Lovelock's theorem, which restricts these deformations to be canonical transformations. In case (II), the deformations of the canonical variables induce higher-derivative terms that increase the number of degrees of freedom. The latter case must be studied carefully since it is possible to alter the theory in a way that does not produce a healthy Hamiltonian representation. We presented an example where the modifications of the metric are traceless and symmetric, which in some sense mimics potential graviton corrections. In this example, we found that the new Hamiltonian in \eqref{eq:Mike-vectortensor hamiltonian density} demands further constraints on the deformations, which are required in order to hold gauge independent scalar and vector constraints. The evaluation of these constraints can only have two possible outcomes: the deformations can either remove the extra degrees of freedom from the generators of the diffeomorphism group, or break some of them by reinterpreting the extra fields as Goldstone modes. In the circumstance of the first case, the dynamics of the additional field has been explored in a more realistic environment \cite{Frolov:2017asg} and it can be benign.

Similarly, we extended this scheme for the tetrad-Palatini and the Ashtekar-Barbero actions. In this case, we suggested the so-called holonomy corrections as a specific form of the deformation (although there is no reason to limit the deformations to just this type). The scenario presented is similar in the sense of the possible introduction of extra degrees of freedom in the system. Nonetheless, if the extra terms appearing in the deformed action (which might or might not include extra fields) are projected into the additional Gauss constraint, the system can be forced to a different metric realization. In the existing literature  \cite{Krasnov:2008zz}, it is possible to find that Lovelock's theorem is by-passed since there is not a unique way to write the theory in these variables. Such a case was not covered in this manuscript. We derived the new constraints and the effects of the deformations in the tetrad variables in the cases when extra degrees of freedom appear (or not) in the system. If the deformations do not alter the shift and lapse gauge orbits, there is no reason to expect a different realization of the constraint algebra, which now will also require closure with the Gauss constraint, and the deformations can be constrained to produce stable gauge fixing.   

\vspace{0.7cm}
\section*{Acknowledgments}
The authors would like to thank Igor Boettcher, J.~Richard Bond, Andrei Frolov, Claire Maulit and Alex Zucca for valuable discussions and correspondence through the realization of this project. JG acknowledges support from the Discovery Grants program of the Natural Sciences and Engineering Research Council of Canada, the Billy Jones Fellowship by the Department of Physics at Simon Fraser University, and by the Perimeter Institute for Theoretical Physics, where some of this work was carried out.  AD acknowledges support from SFU OF.



\newpage 

\linespread{0.6}
\bibliographystyle{unsrt}

} 
\end{document}